\documentclass[a4paper]{article}

\usepackage{nicematrix}
\usepackage{mathtools,amssymb}
\usepackage{physics}
\usepackage{float,subfig}
\usepackage{siunitx}
\usepackage{algorithm}
\usepackage{algpseudocode}
\usepackage{geometry}
\usepackage{tikz,relsize}
\usetikzlibrary{arrows.meta}
\usetikzlibrary{calc}
\usetikzlibrary{math}
\usetikzlibrary{matrix,positioning,arrows.meta,arrows}
\tikzset{fontsize/.style = {font=\relsize{#1}}}

\usepackage{comment}
\usepackage{parcolumns}

\usepackage{xcolor}

\usepackage{hyperref}
\hypersetup{
  colorlinks = true,
  urlcolor   = blue,
  linkcolor  = black,
  citecolor  = black
}

\usepackage[
    backend=biber,
    style=apa,
    natbib=true,
]{biblatex}
\let\cite\parencite
\usepackage{mathtools,amssymb}
\usepackage{physics}
\usepackage{bm}
\newcommand{\fpart}[2]{\partial_{#2}\left(#1\right)}
\newcommand{\fparti}[2]{\partial_{#2}#1}

\makeatletter
\newcommand*{\defeq}{\mathrel{\rlap{%
                     \raisebox{0.3ex}{$\m@th\cdot$}}%
                     \raisebox{-0.3ex}{$\m@th\cdot$}}%
                     =}
\makeatother

\DeclareMathOperator{\diag}{diag}

\newcommand{\patm}{p_{\text{atm}}}

\newcommand{\gradh}{\nabla_{h}}

\newcommand{\uuu}{\bm{u}}
\newcommand{\Transp}{\bm{Q}}

\newcommand{\hpg}{\bm{r}}

\newcommand{\fddde}{\bm{F}_{3D}^\text{h}}
\newcommand{\fdddi}{\bm{F}_{3D}^\text{v}}
\newcommand{\fdd  }{\bm{F}_{2D}}
\newcommand{\fdddh}{\bm{F}_{3D\ra2D}^\text{h}}

\newcommand{\prismh}{\ensuremath{\hat{\mathcal{P}}}}
\newcommand{\gtoph}{\hat{\mathcal{T}}_\text{top}}
\newcommand{\gboth}{\hat{\mathcal{T}}_\text{bot}}

\newcommand{\xx}{\bm{x}}

\newcommand{\ra}{\ensuremath{\rightarrow}}

\newcommand{\R}{\mathbb{R}}

\newcommand{\eint}[1]{\expval{#1}}
\makeatletter
\newsavebox{\@bra}
\newsavebox{\@brb}
\DeclarePairedDelimiterX\myinnerp[1]{.}{.}{%
  \delimsize\langle%
  \hspace*{0.55mm}\savebox{\@bra}{\(\displaystyle\left\langle\vphantom{#1}\right.\)}\hspace*{-0.85\wd\@bra}%
  \delimsize\langle%
  #1% 
  \delimsize\rangle%
  \hspace*{0.55mm}\savebox{\@brb}{\(\displaystyle\left.\vphantom{#1}\right\rangle\)}\hspace*{-0.85\wd\@brb}%
  \delimsize\rangle
}
\makeatother
\newcommand{\eeint}[1]{\myinnerp*{#1}}

\usepackage{stmaryrd}

\def\lllceil{\left\lceil\kern-3.5pt\left\lceil}
\def\rrrceil{\right\rceil\kern-3.5pt\right\rceil}

\newcommand{\inte}{\ensuremath{^\text{int}}}
\newcommand{\exte}{\ensuremath{^\text{ext}}}

\renewcommand{\phi}{\varphi}

\newcommand{\idxh}{_\text{h}}

\newcommand{\jj}{J_{h}} % 2D Jacobian
\newcommand{\nzh}{\hat{n}_z}
\newcommand{\transp}{\bm{q}}

\newcommand{\dfloor}[1]{\left\lfloor #1 \right\rfloor}

\newcommand{\tinybulletblock}{%
  \begin{tikzpicture}[scale=0.2]
    \foreach \x in {0,1,2,3,4,5,6}{
      \foreach \y in {0,1,2}{
        \fill (\x, \y) circle (1.5mm);
      }
    }
  \end{tikzpicture}%
}

\usepackage{afterpage}

\usepackage{xstring}
\title{An efficient multi-GPU implementation for the Discontinuous Galerkin ocean model SLIM}
\date{}

\usepackage[blocks]{authblk}

\author[1]{M. De Le Court}
\author[1]{V. Legat}
\author[3]{A. P. Ishimwe}
\author[2]{C. Scherpereel}
\author[1,2]{E. Hanert}
\author[1]{J. Lambrechts}

\affil[1]{Institute of Mechanics, Materials and Civil Engineering, UCLouvain, Louvain-la-Neuve, Belgium}
\affil[2]{Earth and Life Institute, UCLouvain, Louvain-la-Neuve, Belgium}
\affil[3]{Department of Ecoscience, Aarhus University, Roskilde, Denmark}

\begin{document}
\clearpage

\maketitle

\begin{abstract}
Unstructured-mesh ocean models are increasingly used for coastal applications due to their ability to represent complex geometries and apply local grid refinement where needed. However, their broader use has been hindered by their high computational cost, particularly for models based on the Discontinuous Galerkin finite element (DG-FE) method, which involves significantly more degrees of freedom than traditional finite volume or continuous finite element approaches. The rapid emergence of GPU-based high-performance computing architectures now offers a pathway to address this limitation, as DG-FE formulations are inherently well suited to massively parallel, element-wise computations. Here, we present a full 3D DG-FE ocean model implementation optimized for both single- and multi-GPU systems, with support for both NVIDIA and AMD architectures. We detail the computational strategies employed to achieve high performance, including memory layout optimization, kernel-level parallelization, and matrix-free solvers for key vertical processes. Benchmark results demonstrate that a single HPC-grade GPU (e.g. NVIDIA A100) delivers performance equivalent to approximately 1500 CPU cores, while replacing a 128-core CPU node with a 4$\times$A100 GPU node yields a speedup of around 50$\times$. Weak-scaling efficiency is maintained up to 1024 GPUs. We further demonstrate the model's capabilities on a real-world application in the Great Barrier Reef, achieving a spatial resolution five times finer than the most accurate existing model while maintaining a physical-to-numerical time ratio of 100. These results highlight how GPU-accelerated DG-FE methods can dramatically advance the capabilities of unstructured-mesh ocean modeling, enabling ultra-high-resolution coastal simulations that were previously infeasible.
\end{abstract}

\newpage

\tableofcontents

\section*{Introduction}

% Focus on UG models
Unstructured-mesh ocean models, once considered a niche approach, have progressively gained prominence and now represent a mainstream modelling approach, particularly for coastal ocean studies. Their growing adoption in coastal applications is largely due to their geometrical flexibility, allowing them to accurately represent intricate bathymetric and topographic features through local mesh refinement, as well as to resolve multiscale processes along the land-sea continuum \cite{deleersnijderMultiscaleModellingCoastal2010,lermusiauxMultiscaleModelingCoastal2013}. Unstructured-mesh ocean models can be based on different numerical methods, including the finite volume (FV) method \cite{chenUnstructuredGridFiniteVolume2003,zhangSeamlessCrossscaleModeling2016,danilovFinitevolumESeaIce2017,kornICONOOceanComponent2022}, the continuous Galerkin finite element (CG-FE) method \cite{umgiesserFiniteElementModel2004,wangFiniteElementSea2014,westerinkBasinChannelScaleUnstructured2008} and the Discontinuous Galerkin finite element (DG-FE) method \cite{karnaBaroclinicDiscontinuousGalerkin2013,karnaThetisCoastalOcean2018}. However, despite their advantages, unstructured-mesh models are often limited by a higher computational cost per degree of freedom than structured-grid models. As a result, they remain less commonly employed for global ocean simulations as compared to structured-grid models \cite{danilovFinitevolumESeaIce2017}.

% Computational costs and cost-reduction strategies
The higher computational demand of unstructured-mesh ocean models, and particularly DG-FE models, necessitates the implementation of cost-reduction strategies to help make them viable for large-scale simulations. The DG-FE formulation combines inter-element fluxes (typical of FV methods) with intra-element calculations (typical of CG-FE methods), hence implying more calculations than these two approaches. Additionally, the data organization on an unstructured mesh with irregular connections between elements requires non-trivial indexing and neighbour-finding algorithms due to its non-systematic structure. Consequently, neighbour relationships lack implicit connectivity information and must be explicitly stored and managed. This explicit storage necessitates more sophisticated data handling and memory management methods, ultimately leading to higher computational costs. On the other hand, various strategies have been proposed to address these computational challenges, including local mesh refinement \cite{danilovResolvingEddiesLocal2015}, adaptive time stepping \cite{dawsonParallelLocalTimestepping2013,senyEfficientParallelImplementation2014}, or mode-splitting procedures \cite{shchepetkinRegionalOceanicModeling2005}. The first two strategies complement each other by focusing computing resources where they are most needed and employing smaller time steps exclusively within regions of high mesh resolution. In contrast, the latter optimizes the use of computational resources by splitting the treatment of barotropic and baroclinic processes, using small explicit timesteps for fast processes while employing a single, larger timestep for slower processes. However, these strategies alone are not sufficient to make the routine application of unstructured-mesh models straightforward, as their inherent computational burden remains significant, particularly for high-resolution or long-duration simulations.

% HPC developments
Over the last 10-15 years, significant improvements in model efficiency have also been facilitated by the evolution of high-performance computing (HPC) infrastructures, particularly the widespread transition from CPU-based to GPU-based architectures. Graphics Processing Units (GPUs), characterized by their substantial computational power per chip, have notably accelerated HPC capabilities, reaching exascale performances \cite{changSimulationsEraExascale2023}. This GPU revolution, supported by the development of programming systems such as CUDA, HIP, or OpenCL \cite{dallyEvolutionGraphicsProcessing2021}, is clearly illustrated by the fact that GPU-based supercomputers now dominate the list of the world's fastest HPC systems. To date, nine of the top ten supercomputers in the TOP500 list \cite{top500cite} are accelerated by either INTEL, AMD or NVIDIA GPUs. However, leveraging GPU performance necessitates a complete redesign of computational codes, representing a non-trivial challenge, especially for mature models. Achieving performance, portability and code maintainability have now become three major conflicting objectives preventing well-established modelling frameworks from evolving as fast as their computing clusters.

% new state of models now
Despite this rapid hardware evolution, many of the most widely used structured-grid ocean models remain predominantly CPU-based. Systems such as NEMO \cite{madecNEMOOceanEngine2022}, HYCOM \cite{bleckOceanicGeneralCirculation2002}, ROMS \cite{shchepetkinRegionalOceanicModeling2005}, and MITgcm \cite{marshallFinitevolumeIncompressibleNavier1997} are still implemented mainly in Fortran and parallelized with MPI or OpenMP. While partial GPU adaptations exist such as for ROMS \cite{panzerHighPerformanceRegional2013}, these are generally limited to specific components and do not deliver full model acceleration.

Recent progress in global circulation modelling, however, illustrates what is possible when GPU architectures are fully embraced. LICOM3-HIP \cite{weiAcceleratingLASGIAP2024} demonstrates efficient multi-GPU scaling for eddy-resolving global simulations, while Veros \cite{hafnerFastCheapTurbulent2021} shows how a structured finite-difference model rewritten in Python/JAX can achieve strong performance gains through just-in-time compilation. Oceananigans.jl \cite{wagnerHighlevelHighresolutionOcean2025}, developed natively in Julia using kernel-fusion techniques that make highly effective use of GPU memory bandwidth, represents a new generation of GPU-first models, reaching 10~SYPD at 8~km resolution on 64 GPUs \cite{silvestriOceananigansjlJuliaLibrary2024}. These examples underscore how models purpose-built for GPUs can outperform legacy codes even at global scales.

% Knowledge gap and the present study
Regional ocean modelling still has comparatively few GPU-ready options. In this paper, we present the latest iteration of the three-dimensional unstructured-mesh coastal ocean model SLIM, fully redesigned and optimized to efficiently utilize multiple GPUs. A light abstraction layer allows switching the target architecture between CPUs or NVIDIA or AMD GPUs using the same source code. The use of the Discontinuous Galerkin method in the SLIM model is especially interesting for the GPU implementation. Compared to other unstructured approaches, the DG method enjoys a high data locality and  arithmetic intensity. This lets it use more of the GPUs' computational power. Its block-based discretization is well suited to their high throughput and parallel capabilities. SLIM runs on triangular meshes extruded vertically into columns of prisms. This vertical structure is excellent for the computational efficiency of the model as the penalty of horizontally unstructured accesses can be significantly mitigated. The other major benefit of vertical extrusion is that the implicit treatment of vertical quantities only couples together nodes of the same column of prisms. Therefore, all columns can be processed in parallel, taking full advantage of the GPU.

The first section describes the model’s DG-FE formulation. Section 2 details the GPU-focused implementation on a single GPU. Section 3 presents the multi-GPU strategy. Section 4 evaluates performance through comprehensive benchmarking. Finally, we demonstrate the model’s scientific capability using a realistic case study of the Great Barrier Reef.

\clearpage
\section{Model}\label{sec:model}
\begin{figure}[htpb]
    \centering
    \subfloat[2D mesh around the Whitsundays islands.]{\includegraphics[height=5cm]{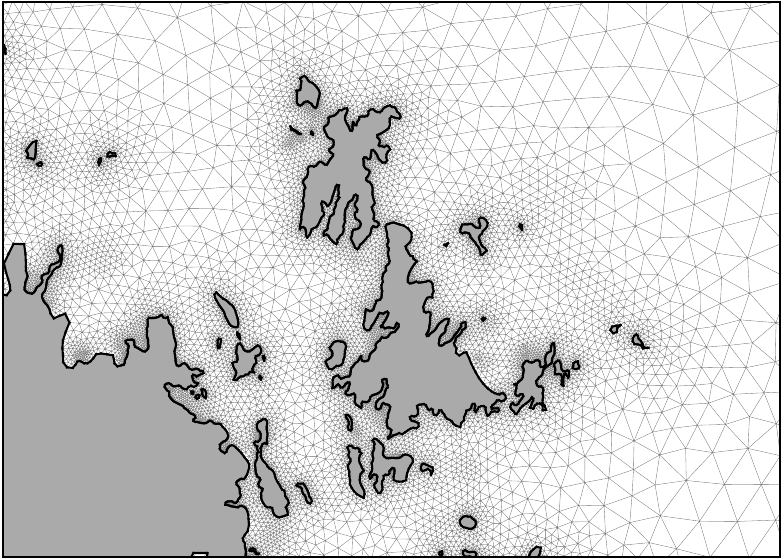}}
    \qquad\qquad\qquad
    \subfloat[Simplified 3D mesh.]{\includegraphics[height=5cm]{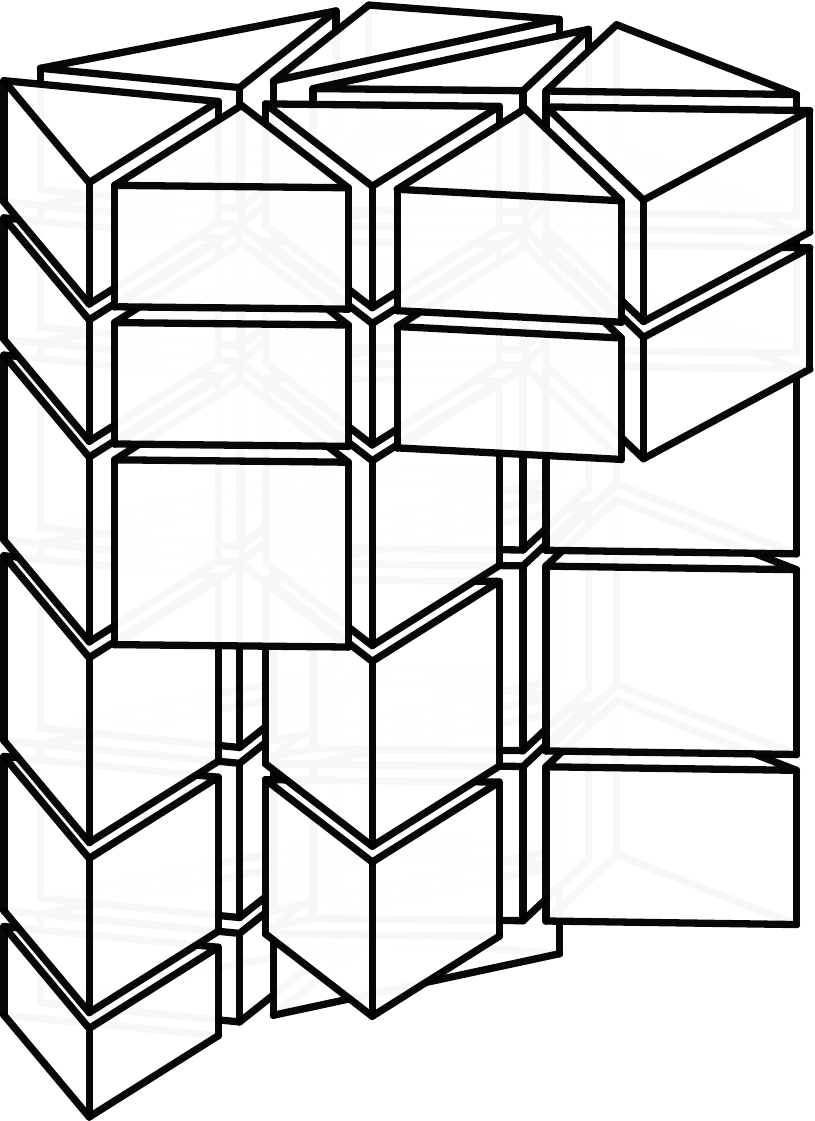}}
    \caption{Examples of meshes used in SLIM. The unstructured 2D triangular mesh is extruded vertically in columns of prisms.}
    \label{fig:mesh:example}
\end{figure}

SLIM is a regional 2D and 3D ocean model. It uses the Discontinuous Galerkin (DG) method with linear Lagrange basis functions. 2D equations are discretized on an unstructured triangular mesh (Figure~\ref{fig:mesh:example}a), while 3D equations use a prismatic mesh made by extruding the 2D mesh vertically (Figure~\ref{fig:mesh:example}b). SLIM 3D solves the hydrostatic Boussinesq equations. This section introduces the numerical formulation of SLIM, which serves as the foundation for the optimizations described in the remainder of the paper. 

\subsection{Primitive equations}
The primitive hydrostatic Boussinesq equations being solved are
\begin{align}
    \fparti{w}{z} =& - \gradh \cdot \uuu \label{eq:conti3d}\\[2mm]
    \fparti{H}{t} = \fparti{\eta}{t} =& - \gradh \cdot \Transp + s \label{eq:mass2d}  \\[2mm]
    \fparti{\uuu}{t} + \gradh \cdot (\uuu \otimes \uuu) + \fpart{w \uuu}{z} =& \nabla \cdot (\bm{\nu} \nabla \uuu)
    - f \bm{e}_z \times \uuu - \dfrac{1}{\rho_0} \gradh p \label{eq:moment3d}
\end{align}    
with the following boundary conditions on the surface and bottom of the ocean:
\begin{align}
    && w - \uuu \cdot \gradh \eta &= \fparti{\eta}{t} - s  & \text{at the ocean surface,} \label{eq:impermeability:surface} \\
    && w - \uuu\cdot \gradh b &= 0 & \text{on the ocean floor.} \label{eq:impermeability:floor}
\end{align}
In the equations above, $H$ denotes the height the water column, $b$ the bathymetry and $\eta$ the deviation from the reference level. $\uuu = [u, v]$ is the 3D horizontal velocity, and $w$ the vertical velocity. $\Transp$ is the horizontal transport defined as
\[
    \Transp = \int_b^\eta \uuu \dd{x}.
\]
The viscosity tensor is written as $\bm{\nu}$, $f$ is the Coriolis parameter and $\bm{e}_z$ is the vertical unit vector. The source term $s$ represents rain ($s > 0$) or evaporation ($s < 0$).
Additionally, temperature and salinity are modelled as tracers, following the advection-diffusion equation
\begin{equation}\label{eq:tracer}
    \fparti{T}{t} + \gradh \cdot (\uuu T) + \fpart{w T}{z} = \nabla \cdot (\bm{\kappa} \nabla T)
\end{equation}
for a generic tracer $T$ and associated diffusivity tensor $\bm{\kappa}$. The parameterization of the eddy viscosity and diffusivity tensors is split between the horizontal and vertical directions.
Vertically, the 2-equation GLS turbulence closure model \cite{umlaufGenericLengthscaleEquation2003} discretized as in \cite{karnaDiscontinuousGalerkinDiscretization2020} is used for the eddy coefficients. Horizontally, the viscosity is parameterized by the Smagorinsky model \cite{smagorinskyGENERALCIRCULATIONEXPERIMENTS1963}, and the diffusivity uses the Okubo model \cite{okuboOceanicDiffusionDiagrams1971}.

\subsection{Split momentum equation}
Under the hydrostatic assumption and accounting for the atmospheric pressure $\patm$, the pressure $p$ can be written in terms of the density $\rho(\xx) = \rho_0 + \rho'(\xx)$ integrated over the water column.
\[
    p(z) = \patm + g \int_{z}^\eta \rho(\tilde{z}) \dd{\tilde{z}} = \patm + g \rho_0 (\eta - z) + g \int_{z}^\eta \rho'(\tilde{z}) \dd{\tilde{z}}
\]
Using this separation, we express the pressure gradient as 
\begin{equation}\label{eq:pressure_gradient}
    \dfrac{1}{\rho_0} \gradh p = \dfrac{1}{\rho_0}\gradh \patm + g \gradh \eta + \dfrac{1}{\rho_0} \hpg
\end{equation}
with the horizontal pressure gradient $\hpg$ obtained by vertical integration
\begin{equation}\label{eq:hpg}
    \begin{cases}
        \hpg = g\rho'(\eta) \gradh \eta \qquad \text{on $\Gamma_s$}  \\[2mm]
        \fparti{\hpg}{z} = g\gradh \rho'
    \end{cases}
\end{equation}
from top to bottom. The momentum equation is thus split as
\begin{align}
    &\fparti{\uuu}{t} + \gradh \cdot (\uuu \otimes \uuu) + \fpart{w \uuu}{z} 
    = \nabla \cdot (\bm{\nu} \nabla \uuu) - f \bm{e}_z \times \uuu - \dfrac{\hpg}{\rho_0} \overbrace{-g \gradh \eta - \dfrac{1}{\rho_0} \gradh \patm}^{S_2} \label{eq:moment3d:split} \\[2mm]
    &\fparti{\Transp}{t} 
    = -g H \gradh \eta - \dfrac{H}{\rho_0} \gradh \patm + \underbrace{\gradh \cdot \int_b^\eta - \uuu \otimes \uuu \dd{z} 
    + \int_b^\eta \nabla \cdot (\bm{\nu} \nabla \uuu) - f \bm{e}_z \times \uuu - \dfrac{\hpg}{\rho_0} \dd{z}}_{S_3} \label{eq:moment2d:split}
\end{align}
where $S_2$ are the 2D terms of the 3D momentum equation and $S_3$ are the inegrated 3D terms in the 2D momentum equation.

In combination with the 2D continuity equation \eqref{eq:mass2d}, the term $gH\gradh\eta$ of the 2D momentum equation \eqref{eq:moment2d:split} creates the dynamics of fast-moving gravity waves. 
Like most ocean models \cite{shchepetkinRegionalOceanicModeling2005,karnaThetisCoastalOcean2018,madecNEMOOceanEngine2022}, SLIM separates the dynamics into a fast barotropic (external) mode and a slower baroclinic (internal) mode.

The barotropic mode represents depth-averaged gravity waves and is integrated on the two-dimensional mesh with a small time step. The baroclinic mode captures the fully three-dimensional internal motions and is advanced with a larger time step. A split-IMEX Runge-Kutta scheme \cite{ishimweSplitexplicitSecondOrder2023,ishimweMultiscaleIMEXSecond2025} couples the two: for each baroclinic step, several barotropic iterations are performed. This mode-splitting approach greatly reduces computational cost compared to applying the small barotropic time step to the full 3D system.
The internal mode is further separated between its vertical and horizontal components. The Horizontal components of the advection, viscosity and diffusion for tracers are computed explicitly, but their vertical counterparts can be implicit. This mitigates the otherwise strict CFL condition arising from the vertical viscosity and diffusion.

More details on the temporal and spatial discretizations can be found in the supporting information.

\clearpage
\section{Efficient GPU implementation of the model}\label{sec:impl_eff}
Having established the mathematical formulation and time-stepping structure, we now turn to the practical question of how these equations can be mapped efficiently to GPUs. We start by giving an overview of SLIM's two-step split-IMEX scheme where the first step is vertically implicit, and the second is explicit. Owing to the mode-splitting approach, each of these steps of the internal mode is accompanied by several iterations of the external mode. The external mode use a three-step explicit Runge-Kutta method.

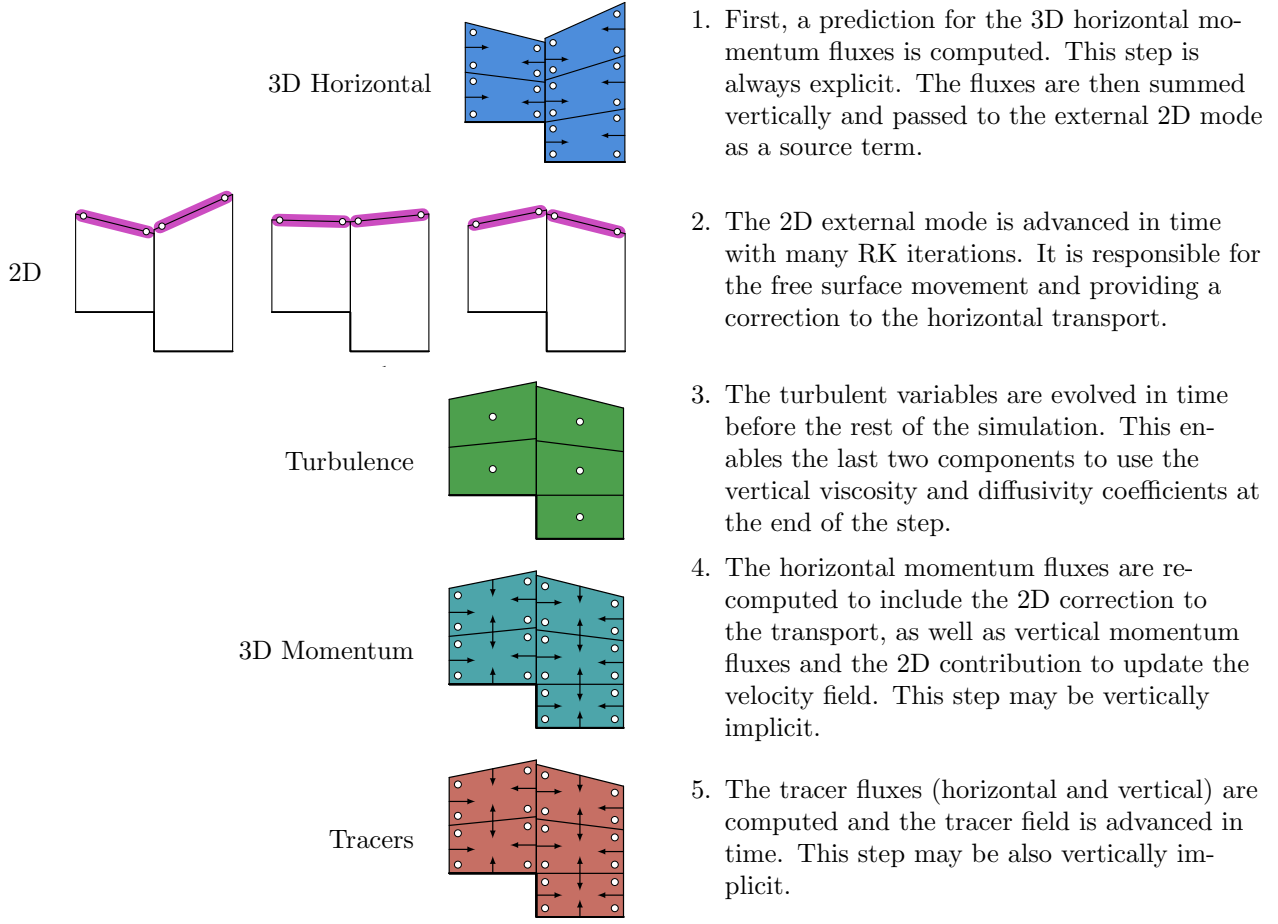
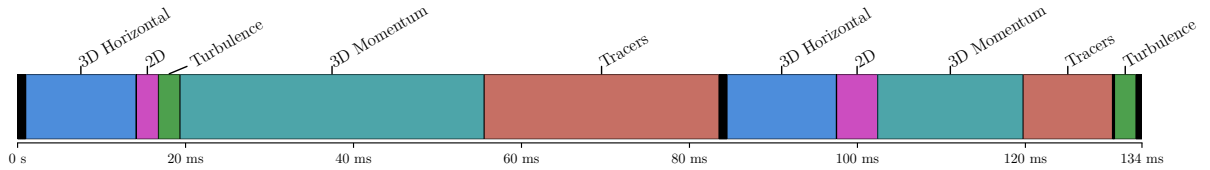
\begin{figure}[htpb]
    \centering
    \subfloat[Schematic view of the five main components of a time step]{
        \makebox[\textwidth][c]{
            \begin{tikzpicture}
                % Include the images and corresponding paragraphs
                \node[left] (img1) at (0,0) {
                    \includegraphics[height=2.5cm]{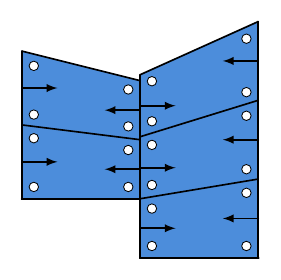}
                };
                \node[right=0cm of img1, text width=9cm] {
                    \begin{minipage}{8cm} \begin{enumerate}\setcounter{enumi}{0}
                        \item First, a prediction for the 3D horizontal momentum fluxes is computed. This step is always explicit. The fluxes are then summed vertically and passed to the external 2D mode as a source term.
                    \end{enumerate} \end{minipage}
                };
                \node[left=0cm of img1] {3D Horizontal};

                \node[left, below=2.5cm of img1.east, anchor=east] (img2) {
                    \includegraphics[height=2.5cm]{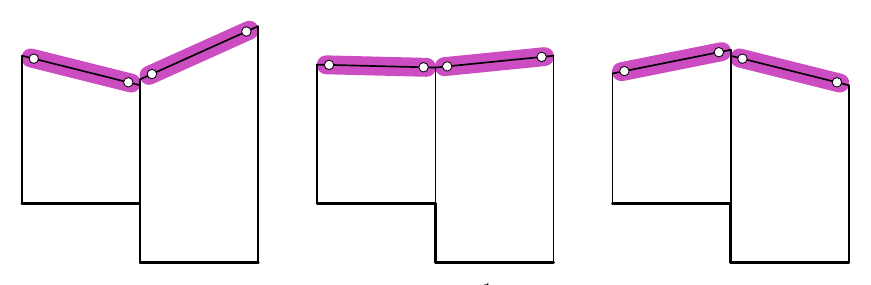}
                };
                \node[right=0cm of img2, text width=9cm] {
                    \begin{minipage}{8cm} \begin{enumerate}\setcounter{enumi}{1}
                        \item The 2D external mode is advanced in time with many RK iterations. It is responsible for the free surface movement and providing a correction to the horizontal transport.
                    \end{enumerate} \end{minipage}
                };
                \node[left=0cm of img2] {2D};

                \node[left, below=2.5cm of img2.east, anchor=east] (img3) {
                    \includegraphics[height=2.5cm]{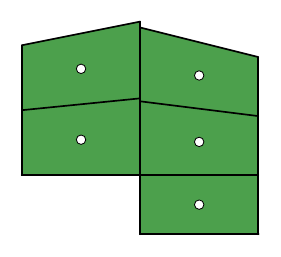}
                };
                \node[right=0cm of img3, text width=9cm] {
                    \begin{minipage}{8cm} \begin{enumerate}\setcounter{enumi}{2}
                        \item The turbulent variables are evolved in time before the rest of the simulation. This enables the last two components to use the vertical viscosity and diffusivity coefficients at the end of the step.
                    \end{enumerate} \end{minipage}
                };
                \node[left=0cm of img3] {Turbulence};

                \node[left, below=2.5cm of img3.east, anchor=east] (img4) {
                    \includegraphics[height=2.5cm]{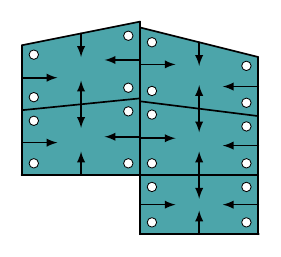}
                };
                \node[right=0cm of img4, text width=9cm] {
                    \begin{minipage}{8cm} \begin{enumerate}\setcounter{enumi}{3}
                        \item The horizontal momentum fluxes are recomputed to include the 2D correction to the transport, as well as vertical momentum fluxes and the 2D contribution to update the velocity field. This step may be vertically implicit.
                    \end{enumerate} \end{minipage}
                };
                \node[left=0cm of img4] {3D Momentum};

                \node[left, below=2.5cm of img4.east, anchor=east] (img5) {
                    \includegraphics[height=2.5cm]{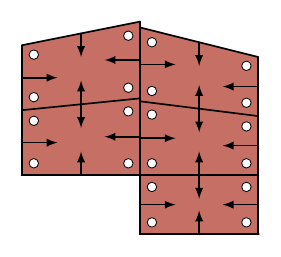}
                };
                \node[right=0cm of img5, text width=9cm] {
                    \begin{minipage}{8cm} \begin{enumerate}\setcounter{enumi}{4}
                        \item The tracer fluxes (horizontal and vertical) are computed and the tracer field is advanced in time. This step may be also vertically implicit.
                    \end{enumerate} \end{minipage}
                };
                \node[left=0cm of img5] {Tracers};

            \end{tikzpicture}
        }%
    }

    \subfloat[Timeline of a full iteration]{
        \includegraphics[width=\textwidth]{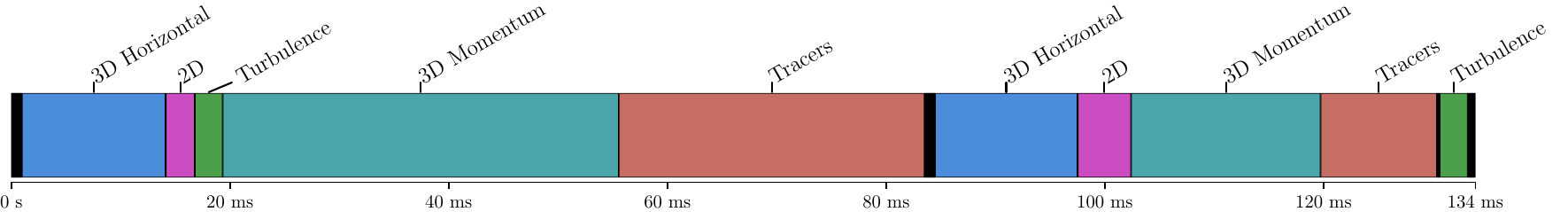}
    }

    \caption{
        (a) Schematic view of the five main components of a time step. The ordering shown corresponds to a vertically implicit step. For vertically explicit steps, the turbulence update is performed last, after the momentum and tracer updates.
        (b) Timeline of a full iteration of the 3D scheme. Black stripes indicate memory operations (e.g., vector initialization). The 3D momentum and tracer components are more expensive in the first step due to the vertically implicit solve. In contrast, the 2D component is more costly in the second step, as it integrates over a larger time interval ($\Delta t$ vs. $\Delta t/2$), leading to more 2D iterations. All timelines in this work were obtained on one or two A100 GPUs, using single precision, on a mesh with 210k triangles and 32 vertical layers. 
    }
    \label{fig:shematic_timestep_first}
\end{figure}

Each step of the internal scheme consists of 5 main components, as illustrated in Figure \ref{fig:shematic_timestep_first}a. Figure \ref{fig:shematic_timestep_first}b presents a timeline of a full iteration, highlighting the 5 components within each step. Those will serve as the guiding structure for the core of this section where we discuss how each compnent is implemented. 

\subsection{Memory layout and thread assignment}\label{sec:memlayout}
In finite element computations, efficiently mapping elements to threads is a crucial consideration. In most kernels, we assign one thread per mesh element, rather than the other common approach of one thread per node. This choice increases the computational workload per thread, which is beneficial for low-order methods like ours, where the GPU often lacks sufficient computations to stay fully utilized, and the number of nodes per element is relatively small.

\begin{figure}[htpb]
    \centering
    \includegraphics[width=0.9\textwidth]{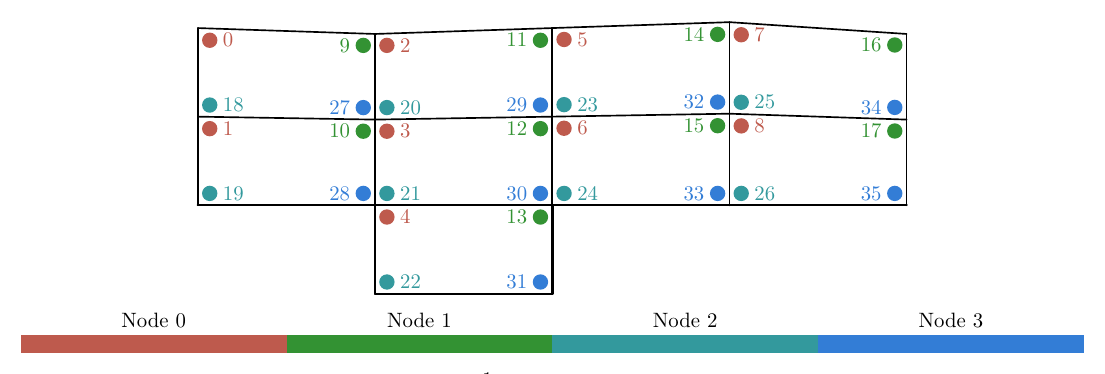}
    \caption{Structure-of-Arrays (SoA) memory layout used in our implementation. Prisms within a column are ordered from top to bottom, with columns arranged sequentially. Nodal values are stored first by node order, followed by element order. For vector fields, each component field is stored contiguously rather than being interleaved. The complete hierarchy is thus field \ra\ node \ra\ column \ra\ layer.}
    \label{fig:memory:soa}
\end{figure}

Given this one-thread-per-element approach, we organize element data in a structure-of-arrays (SoA) format to optimize memory coalescence (Figure \ref{fig:memory:soa}). This differs from the conventional array-of-structures format, where data for each element is stored contiguously. However, the structure-of-arrays approach comes with a drawback: reduced cache efficiency, particularly when dealing with an unstructured mesh. To address this, we reorder the 2D mesh following a Hilbert curve. The 3D mesh, in contrast, is less affected by this issue due to its vertical structure. Neighboring accesses either target the top or bottom elements (likely already in cache) or lateral elements, which are stored in contiguous memory. Consequently, as the number of vertical layers increases, the performance gap between structured and horizontally unstructured models narrows, a property we aim to leverage.
% faire la figure, dire que c'est bien 4 accès en lecture coalescente

\subsubsection{Cell structure for linear system}\label{sec:cell}
The primary drawback of the SoA data layout arises when solving the linear systems associated with each column of prisms. The sequential nature of solving a banded system would ideally suit a one-thread-per-column approach rather than the one-thread-per-prism strategy. With the SoA layout, using one thread per column would result in large strides when accessing memory, leading to poor cache utilization and performance degradation. Distributing memory loads across multiple threads would add synchronization overheads and complexity.

\begin{figure}[htpb]
    \centering
    \includegraphics[width=0.6\textwidth]{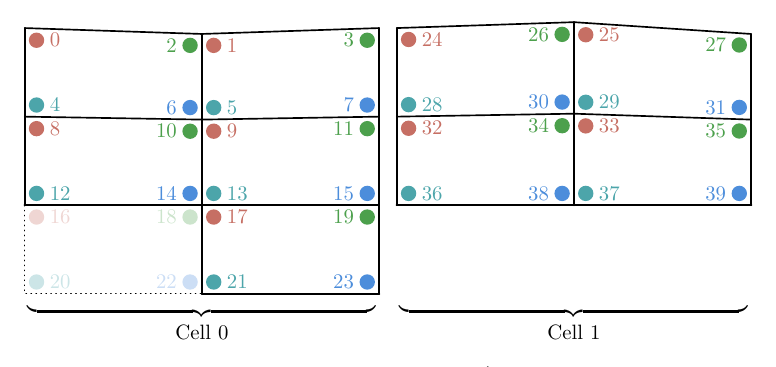}
    \caption{Example of a cell layout for the same mesh as Figure~\ref{fig:memory:soa}. In this minimal example, a cell is made of two columns instead of the typical 128. This figure shows the ordering of the nodes in memory and corresponds to the memory layout of a scalar field. Vector fields are also possible with the complete memory hierarchy being cell \ra\ layer \ra\ node \ra\ field \ra\ column.}
    \label{fig:memory:cell}
\end{figure}

To address this, we maintain a single-thread-per-column approach but introduce what we call the cell layout (Figure~\ref{fig:memory:cell}), a data structure designed to optimize memory access patterns for linear solves while enabling efficient conversion to and from the SoA layout.
\begin{figure}[htpb]
    \centering
    \includegraphics{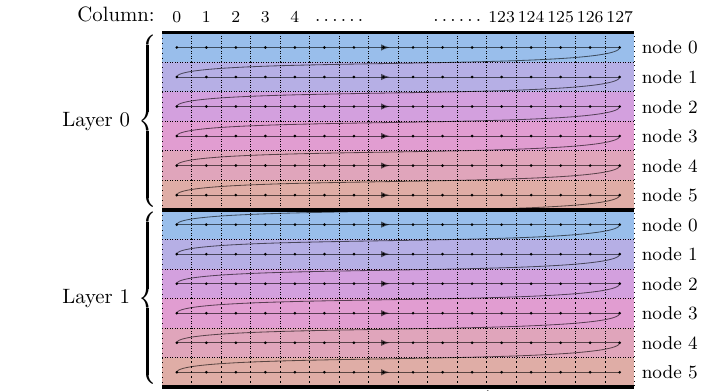}
    \caption{Example of a cell matrix with two layers and 128 columns for a scalar field (6 values per layer). The arrows indicate the order in which the data is stored in memory.} 
    \label{fig:cell_matrix}
\end{figure}
The cell layout groups sets of prism columns (typically 128) and stores their data in a matrix (Figure~\ref{fig:cell_matrix}). In the following, the term \emph{cell} refers to such a group of 128 columns. Each column of the matrix correspond to one column of prisms. Each row corresponds to a data entry from the column of prisms. For example, a scalar field would be unrolled to $6N_\text{layers}$ rows, 6 being the number of nodes per prism. When the number of layers varies, padding is added to match the deepest column of the cell. This ensures that a block of 128 threads solving the linear systems for 128 columns of prisms achieves perfectly coalesced memory access. The layout is akin to an array-of-structure-of-array (AoSoA), which is well suited for GPU computing.

\subsubsection{Going to and from the cell layout}\label{sec:thread_assignment}
We assemble the system with one thread per element. A cell is divided into multiple GPU blocks of threads, each responsible for a subset of the cell's columns. If a block of threads handles $n$ columns, it writes to the cell in contiguous chunks of $n$ values and reads from the SoA layout in chunks of $\dfloor{128/n}$ values.  
\begin{figure}[htpb]
    \centering
    \includegraphics{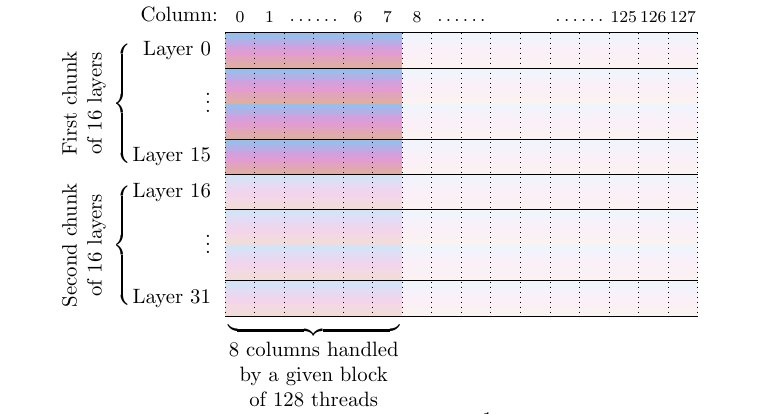}
    \caption{Subset of a cell processed by a block of 128 threads. This example shows a cell with 128 columns and 32 layers being processed by blocks of 128 threads. Here, the block reads from the SoA layout in chunks of 16 values and writes to the cell in chunks of 8 values. Since the cell has 32 layers, the block processes its assigned columns in two passes: first for layers 0 to 15, then for layers 16 to 31.}
    \label{fig:block_per_cell}
\end{figure}
The choice of $n$ is a trade-off. A larger $n$ improves write efficiency to the cell and benefits kernels assembling larger matrices. Conversely, a smaller $n$ enhances coalescence when reading data and is preferable when no linear system needs to be solved. In practice, $n$ is determined per cell based on the maximum number of layers in the cell and an estimate of the average execution time across all kernels.  

In our case, prioritizing vertical size ($\dfloor{128/n}$) over horizontal size ($n$) is generally more effective, as most kernels perform more reads than writes, and many do not require solving a system. With 128 threads per block, a typical choice is $n=8$, meaning each block processes chunks of $8$ columns $\times$ $16$ layers at a time (Figure~\ref{fig:block_per_cell}). If a column has more than $16$ layers, the block iterates over the remaining layers in steps of $16$, and so on.

After assembling a matrix and its right-hand side (RHS) in the cell format, the system is solved while maintaining the same layout. Since all columns are independent, the solving step remains parallelizable over the individual columns.

Finally, a transposition step is required to convert the results back to the SoA layout. This is performed by a kernel with the same structure as the assembly kernels. Thanks to the block-based nature of reads and writes between the cell and SoA layouts, this kernel nearly achieves peak memory bandwidth. Any imperfections in access sizes and alignment are fully masked by the cache hierarchy.

\subsection{3D hydrodynamical scheme}\label{sec:3d_hydro_scheme}
The tracer and the hydrodynamics schemes are very similar in structure, thus we will only detail the hydrodynamics scheme here. From a high level, a step of the hydrodynamics scheme involves solving the following equation:
\begin{equation}\label{eq:moment3d:discrete:simple}
    \dfrac{M_1 \uuu_1 - M_0 \uuu_0}{\Delta t} = \fddde(\uuu, \bar{\transp}, \hpg) + M_1(\fdd/H_1) + \fdddi(\uuu, \tilde{w}, \uuu_1)
\end{equation}
where the unknown is $\uuu_1$, the velocity field at the end of the step, $\uuu_0$ is the velocity field at the start of the step, $\uuu$ is current the velocity field from the Runge-Kutta scheme and $\bar{\transp}$ is the horizontal transport used to advect $\uuu$. $\fddde$ represents the discrete horizontal advection and viscosity fluxes, $\fdd$ denotes the horizontal momentum input from the external 2D mode, and the vertical fluxes are given by $\fdddi(\uuu, \tilde{w}, \uuu_1)$. $M_0$ and $M_1$ are the mass matrices at the beginning and end of the step. The implicit vertical fluxes have been linearized such that $\fdddi(\uuu, \tilde{w}, \uuu_1) = A(\uuu, \tilde{w})\, \uuu_1$. Using that property, equation \eqref{eq:moment3d:discrete:simple} can be rewritten as
\begin{equation}\label{eq:moment3d:discrete:system}
    (M_1 - \Delta t\, A(\uuu, \tilde{w}))\,\uuu_1  = M_0 \uuu_0 + \Delta t\, \big(\fddde(\uuu, \bar{\transp}, \hpg) + M_1(\fdd/H_1)\big)
\end{equation}
where the linear system to be solved appears explicitly. We will now focus on each of these three contributions ($\fddde$, $\fdd$ and $\fdddi$ (or $A$)), and the kernels involved in their computation.

\subsubsection*{3D horizontal flux prediction}
As detailed in Section~S3.2 of the supporting information, the horizontal fluxes $\fddde$ are computed twice. Once with a transport $\transp$ as a prediction step, and later with a transport $\bar{\transp}$ to advance the momentum in time. The prediction of $\fddde$ involves two main steps. First, the density is updated and used to compute the hydrostatic horizontal pressure gradient $\hpg$. Then, $\hpg$ is used to evaluate $\fddde(\uuu, \transp, \hpg)$. A detailed derivation of the discretization of these terms can be found in Sections S2.2 and S3.2 of the supporting information.
\begin{figure}[htpb]
    \centering
    \includegraphics[width=\textwidth]{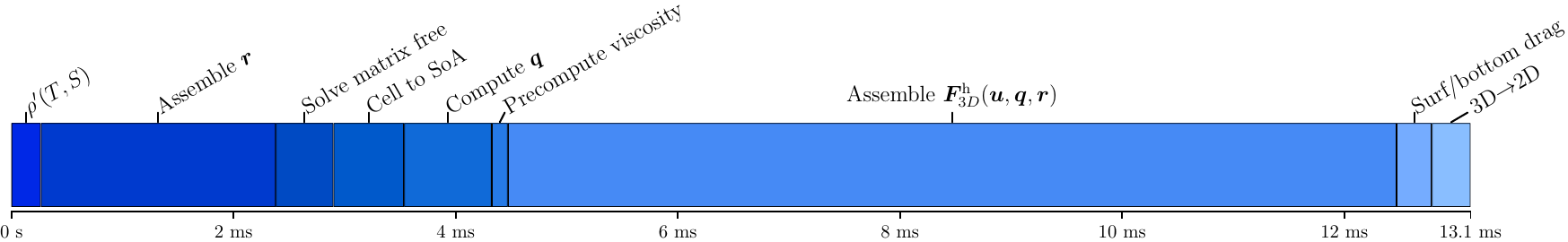}
    \caption{Kernels used for the computation of the horizontal terms of the momentum equation.}
    \label{fig:overview_u_hor}
\end{figure}
The complete timeline for the assembly of the horizontal terms is illustrated in figure \ref{fig:overview_u_hor}. The dominant kernels are as expected the assembly of the RHS for $\hpg$ and the computation of $\fddde$. Computing the horizontal pressure gradient $\hpg$ requires the solution of a linear system. Fortunately, the matrix has a peculiar structure (See Section~\ref{sec:matrix_free_solvers}) that allows solving the system without ever assembling it, hence why the solving step is so short. Computing the horizontal transport $\transp$ and the horizontal viscosity fields are operations that could be integrated in the assembly of $\fddde$, but it is both simpler and faster to precomputed them at all nodes beforehand. The two small kernels at the end are the surface and bottom drag, and vertically summing $\fddde$ to produce $\fdddh$ for the coupling with the external mode. The reason for both of these is explained Section~S3.2 of the supporting information, but we won't go over the details since they remain negligible in the timings.

\subsubsection*{Fluxes from the 2D external mode}
\begin{figure}[htpb]
    \centering
    \includegraphics[width=\textwidth]{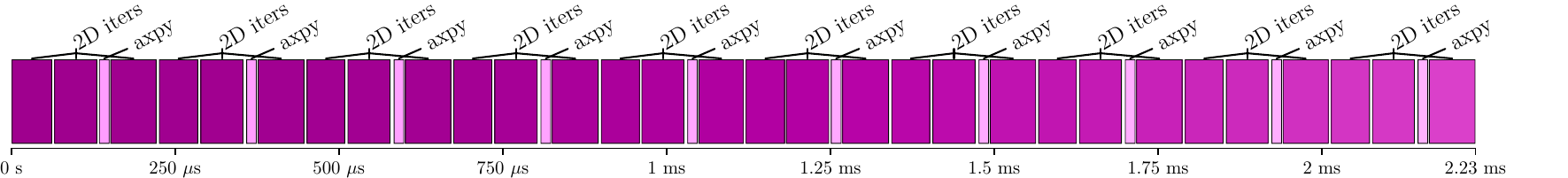}
    \caption{Kernels of the 2D external mode during a full step of the scheme. In this example, the external mode performs 10 iterations of a 3-stage Runge-Kutta scheme.}
    \label{fig:overview_u_2d}
\end{figure}

The external mode produces the fluxes $\fdd$ due to the gravity wave, and also accumulate the mean transport $\bar{\Transp}$ during all the 2D iterations. The momentum change $\fdd$ is computed based on the difference of 2D transport before and after 2D iterations, as detailed in Section~1.2 of the supporting information.
\[
    \fdd = (\Transp_1 - (\Transp_0 + \Delta t \fdddh)) / \Delta t
\]
In order to get a conservative and consistent advection scheme, the vertical sum of the 3D transport $\bar{\transp}$ must exactly match the averaged 2D transport $\bar{\Transp}$. Therefore, at the end of each Runge-Kutta step of the 2D mode, a kernel adds the current solution to the build the time-averaged transport $\bar{\Transp}$. This explains the 3+1 kernels for each iteration of the 2D mode illustrated in Figure~\ref{fig:overview_u_2d}. 

\subsubsection*{3D momentum update with implicit time-stepping}
\begin{figure}[htpb]
    \centering
    \includegraphics[width=\textwidth]{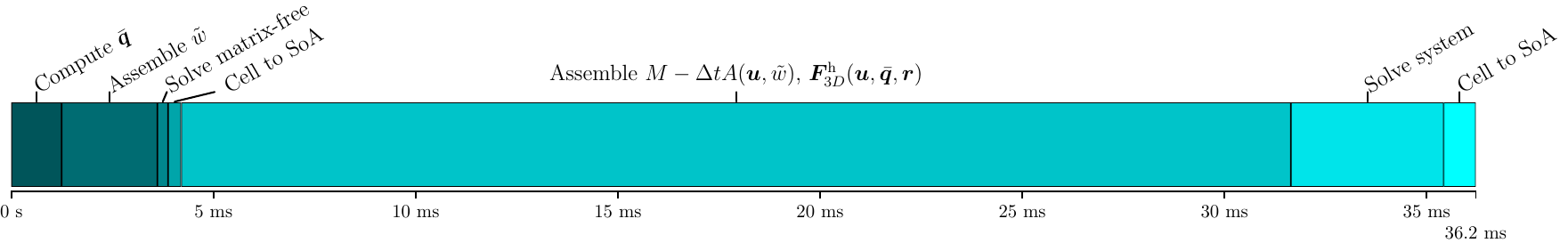}
    \caption{Kernels used for the computation of the vertical terms of the momentum equation during an implicit timestep.}
    \label{fig:overview_u_vert}
\end{figure}
With implicit time-stepping, the momentum update is divided into four steps. First, a corrected transport $\bar{\transp}$ is computed from the vertically averaged 2D transport $\bar{\Transp}$ and the velocity $\uuu$. This transport is then used to compute the vertical velocity $\tilde{w}$. Solving for $\tilde{w}$ requires the solution of a linear system with a structure similar to that of the horizontal pressure gradient $\hpg$, which explains the short solution times observed in Figure~\ref{fig:overview_u_vert}. Next, a single kernel performs the assembly of both the matrix $M - \Delta t\, A(\uuu, \tilde{w})$ and the corresponding right-hand side. The latter involves recomputing $\fddde(\uuu, \bar{\transp}, \hpg)$ using the corrected transport $\bar{\transp}$, and adding the contributions from $M_0 \uuu_0$ and the 2D terms. Finally, the resulting linear system is solved. In contrast to the system for $\tilde{w}$, this system has no exploitable structure beyond its banded form, leading to the longer assembly and solution times shown in Figure~\ref{fig:overview_u_vert}.

\subsubsection*{3D momentum update with explicit time-stepping}
During fully explicit substeps, the temporal scheme remains almost identical. The only difference is that $\fdddi$ no longer depends on $\uuu_1$, so equation \eqref{eq:moment3d:discrete:simple} becomes

\begin{equation}\label{eq:moment3d:discrete:simple_expl}
    \dfrac{M_1 \uuu_1 - M_0 \uuu_0}{\Delta t} = \fddde(\uuu, \hpg) + M_1(\fdd/H_1) + \fdddi(\uuu, \tilde{w}, \uuu)
\end{equation}
or equivalently
\begin{equation}\label{eq:moment3d:discrete:system_expl}
    \uuu_1  = M_1^{-1}\left(M_0 \uuu_0 + \Delta t\, \big(\fddde(\uuu, \hpg) + M_1(\fdd/H_1) + \fdddi(\uuu, \tilde{w}, \uuu)\big)\right) \; .
\end{equation}
Since the mass matrix $M_1$ is block-diagonal, solving a full linear system per column is no longer required, and the values of $\uuu_1$ can be computed independently for each element.
\begin{figure}[htpb]
    \centering
    \includegraphics[width=\textwidth]{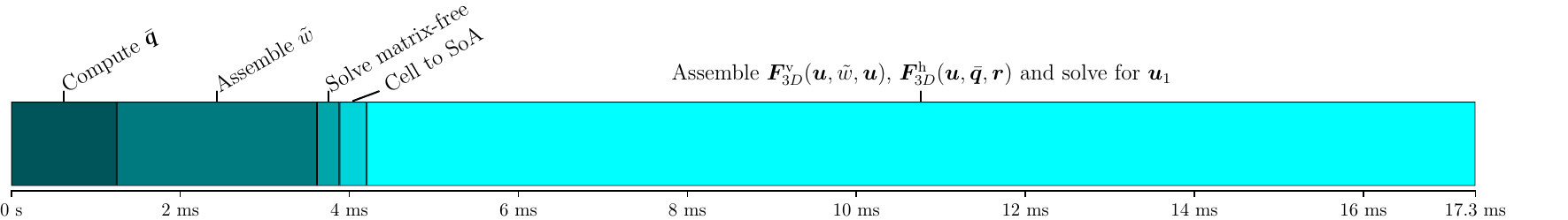}
    \caption{Kernels used for the computation of the vertical terms of the momentum equation during an explicit step.}
    \label{fig:overview_u_vert_expl}
\end{figure}
Avoiding matrix assembly and eliminating the need to write it to global memory frees up significant computational resources and memory bandwidth. This is why the explicit step is considerably faster than the implicit one.

\subsection{Matrix-free solvers}\label{sec:matrix_free_solvers}
The equations for the horizontal pressure gradient $\hpg$ and the vertical velocity $w$ share a common structure:
\[
    \dfrac{\partial (\bullet)}{\partial z} = f \;.
\]
In such cases, the discrete equations simplify significantly, allowing the resulting linear systems to be solved without explicitly assembling the matrix.

\subsubsection*{Horizontal pressure gradient}
The discrete form of the horizontal pressure gradient \eqref{eq:hpg} is detailed in Section~S2.2 of the supporting information and essentially takes the form
\begin{equation}\label{eq:hpg:discrete:simple}
    \eeint{\phi \hpg\exte \jj}_{\gtoph} + \eeint{\phi \hpg\inte \jj}_{\gboth} - \eint{\hpg\fpart{\phi}{\hat{z}} \jj}_{\prismh} = \bm{F} \;.
\end{equation}
The left-hand side of this equation couples the degrees of freedom within each vertical column, resulting in one linear system per column. Consequently, the computation of $\hpg$ is divided into two stages: the assembly of the right-hand side of equation \eqref{eq:hpg:discrete:simple}, and solving the system. 

Due to the tensor product formulation of the mesh (i.e. the 3D mesh is made by extruding the 2D mesh in layers of prisms), all the terms that depend on the vertical coordinate cancel out and the system on each column can be written as 
\begin{equation}
    D_{vu} \hpg = \bm{F}
\end{equation}
with $D_{vu}$ a constant matrix per column that only depends on the 2D mass matrix $M_h$. For example, on a 3-layer column we get
\[
    D_{vu} = \dfrac{1}{2} \begin{bNiceArray}{cc:cc:cc}
        -M\idxh & -M\idxh &         &         &         &          \\
         M\idxh & -M\idxh &         &         &         &          \\ \hdottedline
                & 2M\idxh & -M\idxh & -M\idxh &         &          \\
                &         &  M\idxh & -M\idxh &         &          \\ \hdottedline
                &         &         & 2M\idxh & -M\idxh & -M\idxh  \\
                &         &         &         &  M\idxh & -M\idxh 
    \end{bNiceArray}
    \quad \text{with} \quad
    M\idxh = \dfrac{\jj}{24}
    \begin{bmatrix}
        2 & 1 & 1 \\
        1 & 2 & 1 \\
        1 & 1 & 2
    \end{bmatrix}
\]
and the pattern continues with more layers. Knowing the structure of the linear system in advance removes the need to assemble an expensive matrix. Solving the system is done via a single-pass up-looking solver that encodes the matrix structure in a recursion formula. This leads to leading to the matrix-free solver described in algorithm \ref{alg:hpg_matrix_free}.
\begin{algorithm}[htbp]
    \caption{Matrix free solver for $\hpg$ on a single column of prisms}
    \label{alg:hpg_matrix_free}
    \begin{algorithmic}
        \State $\hpg \gets \text{RHS}$ \Comment{Initialize $\hpg$ to solve in-place}
        \State $M \gets \diag(M_h) \in \R^{6\times6}$ \Comment{Bloc-diagonal mass matrix}
        \State $s \gets \bm{0} \in \R^{3\times2}$ \Comment{Initialize local accumulator}
        \State $r \in \R^{6\times2}$ \Comment{Allocate the local $\hpg$}
        % \State $M \gets \begin{bmatrix}M_h & 0\\ 0 & M_h\end{bmatrix} \in \R^{6\times6}$

        \For{$l = 1, \dots, N_\text{layers}$} \Comment{Loop over all layers}
            \State $r \gets \Call{LoadLayer}{\hpg, l}$  \Comment{Load layer in local memory}
            \State $r \gets M^{-1} r$ \Comment{Inverse the 2D mass matrix}
            \For{$i = 1, \dots, 3$} \Comment{Loop over the 3 nodes of a face}
                    \State $s_{i} \gets s_{i} + r_{i} + r_{i+3}$ \Comment{Update the accumulator}
                    \State $r_{i} \gets - s_{i} + 2 r_{i+3}$     \Comment{Compute $\hpg$ on the top face}
                    \State $r_{i+3} \gets -s_{i}$                \Comment{Compute $\hpg$ on the bottom face}
            \EndFor
            \State $r \gets \Call{WriteToLayer}{\hpg, r, l}$     \Comment{Write the local result to global memory}
        \EndFor
    \end{algorithmic}
\end{algorithm}
Since the RHS of the system is stored and solved in the cell format, as described in Section~\ref{sec:cell}, an additional step is required after solving to convert the solution to the AoS layout.

\subsubsection*{Vertical velocity}
Similarly, the equation for the vertical velocity~\eqref{eq:conti3d}, whose discretization is described in Section~S2.1 of the supporting information, takes the form
\[
    \eeint{\phi w\inte \nzh \jj}_{\gtoph} + \eeint{\phi w\exte \nzh \jj}_{\gboth} - \eint{w \jj \fparti{\phi}{\hat{z}}}_{\prismh} = F \;.
\]
As with $\hpg$, the resulting linear system for each vertical column can be written as
\begin{equation}
    D_{vd} w = F \;.
\end{equation}
For a 3-layer column, for instance, the matrix $D_{vd}$ takes the form:
\[
    D_{vd} = \dfrac{1}{2} \begin{bNiceArray}{cc:cc:cc}
         M\idxh & -M\idxh &         &         &         &          \\
         M\idxh &  M\idxh &-2M\idxh &         &         &          \\ \hdottedline
                &         &  M\idxh & -M\idxh &         &          \\
                &         &  M\idxh &  M\idxh &-2M\idxh &          \\ \hdottedline
                &         &         &         &  M\idxh & -M\idxh  \\
                &         &         &         &  M\idxh &  M\idxh 
    \end{bNiceArray}
\]
using the same 2D mass matrix $M\idxh$ as in the $\hpg$ system. Thanks to the predictable structure of $D_{vd}$, this system can also be solved using a matrix-free solver analogous to the one described in Algorithm~\ref{alg:hpg_matrix_free}.

\subsection{Fully-assembled column solvers}\label{sec:matrix_full_solvers}
Beyond the equations for $\hpg$ and $w$, where matrix-free solvers are applicable, other parts of the model require solving linear systems that couple all nodes within a column and cannot rely on a-priori know matrices. This is the case in the turbulence closure model, and in the computation of vertical fluxes during the vertically implicit substeps for both tracers and hydrodynamics.

The matrices involved in the momentum equation~\eqref{eq:moment3d:discrete:system} and the tracer equation share a common sparsity structure. Each element has six nodes, with the top three nodes coupled to the six nodes of the element above, and the bottom three nodes coupled to the six nodes of the element below. For a 3-layer column of prisms, this results in the following sparsity pattern:
\[
    M - \Delta t A(\uuu,w) = 
    \begin{bNiceArray}{c|c|c}
    \\[-3.5mm]
    \tinybulletblock &                  &                  \\
    \tinybulletblock & \tinybulletblock &                  \\ \hline \\[-3.5mm]
    \tinybulletblock & \tinybulletblock &                  \\
                    & \tinybulletblock & \tinybulletblock \\ \hline \\[-3.5mm]
                    & \tinybulletblock & \tinybulletblock \\
                    &                  & \tinybulletblock
    \end{bNiceArray}
\]
Because memory access is expensive, only the nonzero structure of the matrix is stored during assembly. The linear systems are then solved in-place using Gaussian elimination, with one thread assigned per system. Leveraging the sparsity pattern, a local buffer of size 36 is sufficient to store the relevant portion of the matrix at any time. The right-hand side only requires 6 values (or 12 for a 2-component vector like $\uuu$), all of which fit within GPU registers. This eliminates register spilling and ensures that global memory bandwidth is used efficiently for meaningful data transfers.

The turbulence closure model is comparatively much simpler. It involves only a single degree of freedom per element, resulting in tridiagonal linear systems. These systems are stored using a diagonal matrix format and are solved via Gaussian elimination.

\subsection{Data management}\label{sec:data}
Data movement is a primary performance bottleneck on GPUs. This includes transfers within device memory and, more critically, transfers from host memory or from disk. As a regional model, SLIM relies on time-dependent external forcing (e.g., boundary currents, temperature, salinity, wind). Updating forcing fields on the CPU and transferring them to the GPU at every timestep is infeasible given the execution rate of GPU kernels, where multiple updates may be required within tens of microseconds. Consequently, forcing data must be updated directly on the device whenever possible.

To this end, forcing fields are chosen to vary linearly in time between two precomputed input states, typically separated by one hour. Temporal interpolation is performed within compute kernels, thereby minimizing data transfers and avoiding additional kernel launches. When a kernel requires data outside the available temporal window, new data are loaded from disk, interpolated in space, and transferred asynchronously to the GPU. Because this loading is performed on demand, short wait times may occur; however, these delays are negligible in practice. Although not currently necessary, data prefetching may become beneficial as dataset sizes increase.

\subsubsection*{Linear transport while advancing the external mode}
As noted in Section~\ref{sec:3d_hydro_scheme}, tracer consistency requires the time-averaged solution of the external (2D) mode over a full step. This average is obtained by accumulating the 2D solution at each external iteration (the \texttt{axpy} kernels in Figure~\ref{fig:overview_u_2d}) and is subsequently used in the 3D tracer kernels. The same requirement applies to external forcing: the tracer must be driven by a velocity corresponding to the time average of the external transport seen by the 2D mode.

The tracer requires a 3D boundary condition, whereas the external mode uses a 2D one. At the discrete level, for a full step in which the external mode performs $m$ iterations, consistency requires
\[
    \dfrac{1}{m}\sum_{i=0}^{m-1} \Transp_d\!\left(t_0+(i+1/2)\Delta t /m\right)
    =
    \sum_{\text{Vertical DOFs}} \transp_d\!\left(t_0+\Delta t/2\right),
\]
where $\Transp_d(t)$ and $\transp_d(t)$ denote the 2D and 3D external transports at time $t$, respectively. A consistent definition would be to set the 2D transport as the vertical sum of the 3D transport,
\begin{equation}\label{eq:transp_vertical_sum}
    \Transp_d(t) \defeq \sum_{\text{Vertical DOFs}} \transp_d(t),
\end{equation}
and to explicitly time-average $\transp_d$ so that the tracer uses
\[
    \transp_{d,\text{tracer}}\!\left(t_0+\Delta t/2\right)
    \defeq
    \dfrac{1}{m}\sum_{i=0}^{m-1} \transp_d\!\left(t_0+(i+1/2)\Delta t /m\right).
\]
However, this approach would require evaluating vertical sums of 3D data at every 2D iteration and storing the accumulated values, which is computationally expensive. Instead, we interpolate $\transp_d$ linearly in time over $[t_0, t_0+\Delta t]$. Under this assumption, $\Transp_d(t_0)$ and $\Transp_d(t_0+\Delta t)$ are computed using~\eqref{eq:transp_vertical_sum}, and $\Transp_d(t)$ is obtained by linear interpolation between these two states. Since $\Delta t$ is much smaller than the temporal resolution of the forcing data, this approximation introduces no significant loss of accuracy.

\clearpage
\section{Multi-GPU}\label{sec:multi_gpu}
While single-GPU performance is critical, large-scale coastal simulations demand the use of many GPUs working in parallel. Multi-GPU parallelism within SLIM is handled through domain decomposition and distributed memory with MPI, with one GPU per MPI rank. The domain is split horizontally, such that each GPU owns a subset of the triangles of the 2D mesh and its associated columns. Each partition also stores a layer of ghosts triangles from the neighboring partitions to be used in the computations that require neighboring accesses. Such computations are then followed by a halo exchange, to update the ghost elements.

\subsection{Structure of a compute kernel}
Since GPUs significantly accelerate computations compared to CPUs, the relative cost and frequency of communications increase accordingly. Overlapping computation and communication therefore becomes essential. To achieve this, the computation within each partition is split into two parts: (i) boundary elements and elements adjacent to ghost cells, and (ii) interior elements. Boundary computations are launched first, and as soon as they complete, halo exchanges are initiated while interior computations continue.

This decomposition also enables kernel specialization. Interior kernels can assume the presence of all neighbors and avoid conditional logic required at boundaries. While this distinction has limited impact on CPUs, it improves GPU performance by reducing control flow divergence and lowering resource usage (e.g., registers and shared memory).

To enable overlap on the GPU, two streams are used. The \emph{compute stream} handles the bulk of the computations, while the \emph{communication stream} is reserved for boundary computations and halo exchange operations. The communication stream is assigned the highest priority. A full iteration proceeds as follows:
\begin{enumerate}
    \item Boundary computations are launched on the communication stream.
    \item Interior computations are launched asynchronously on the compute stream.
    \item A packing kernel is launched on the communication stream to gather data for halo exchange, as the memory layout is not contiguous per element.
    \item The CPU synchronizes with the communication stream to ensure packing completion, then initiates MPI communications.
    \item Upon completion of MPI communications, an unpacking kernel is launched on the communication stream.
\end{enumerate}
GPU events are then used to ensure that the compute stream waits for communication to complete before proceeding to the next step.

Since MPI predates widespread GPU usage, not all implementations are GPU-aware, i.e., capable of directly accessing device memory. In such cases, halo exchanges require additional copies between host and device memory.

\subsection{MPI communications for the 3D kernel}
\begin{figure}[htpb]
    \centering
    \includegraphics[width=\textwidth]{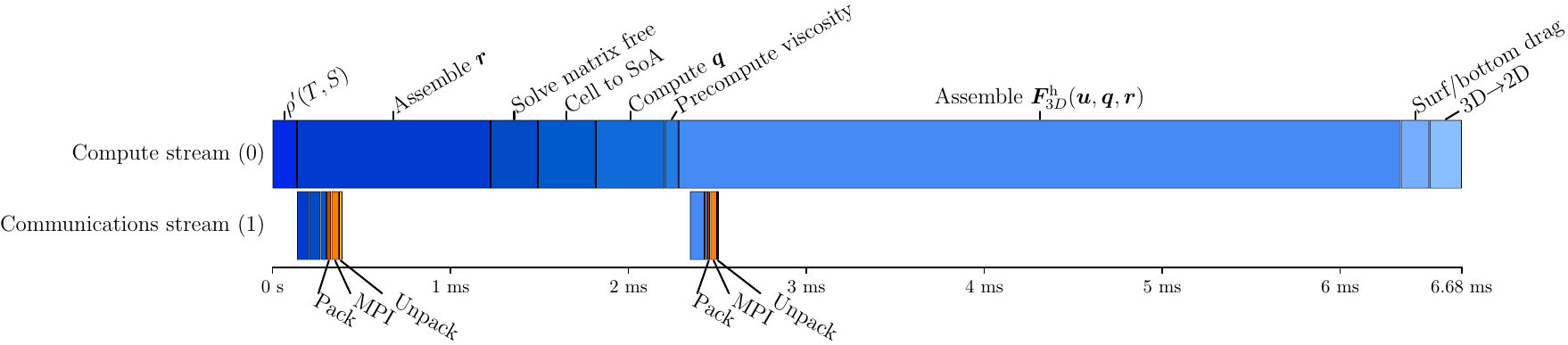}
    \vspace{2mm}
    \includegraphics[width=\textwidth]{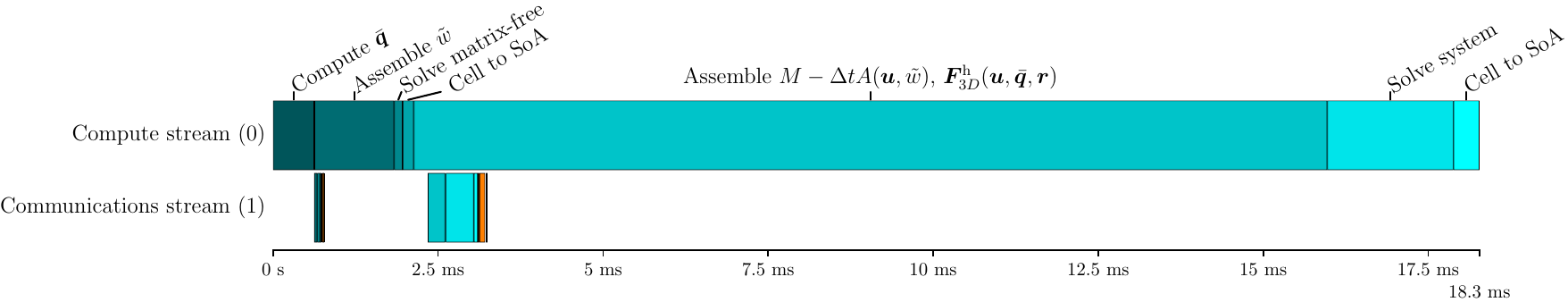}
    \vspace{2mm}
    \includegraphics[width=\textwidth]{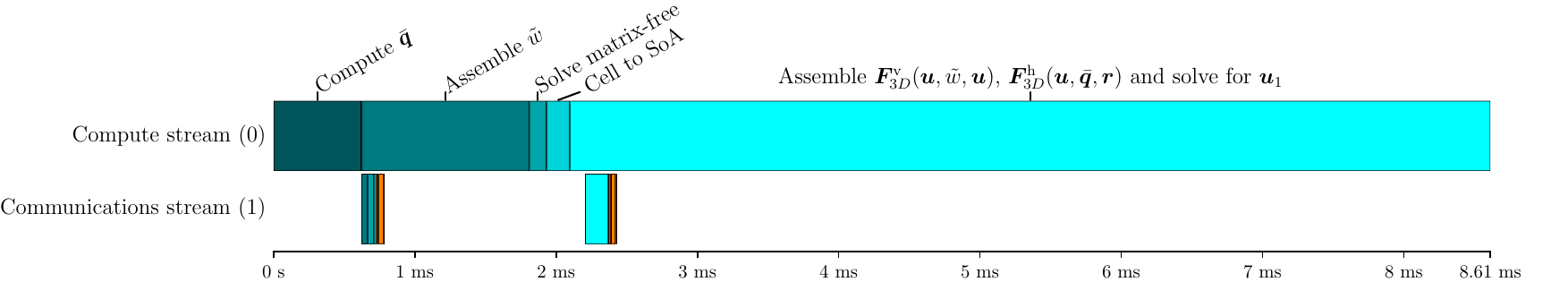}
    \caption{Timeline of both the Compute stream and Communications stream for the two main phases of internal mode of the hydrodynamics computations. On top are the horizontal fluxes, in the middle the vertical fluxes during an implicit step, and on the bottom the vertical fluxes during an explicit step.}
    \label{fig:overview_u_mpi}
\end{figure}
The 3D kernels, as shown in Figures \ref{fig:overview_u_hor}, \ref{fig:overview_u_vert} and \ref{fig:overview_u_vert_expl}, are quite computationally heavy, and are thus not bound by latency. 
As a result, the communications easily overlap with the rest of the computations and the GPU is kept busy, as seen in Figure~\ref{fig:overview_u_mpi}. Compute kernels taking much longer than the synchronous MPI communications also means that the next kernels will already been enqueued by the time the computation is over.

\subsection{Challenges with the short 2D kernels}
\begin{figure}[htpb]
    \centering
    \includegraphics[width=\textwidth]{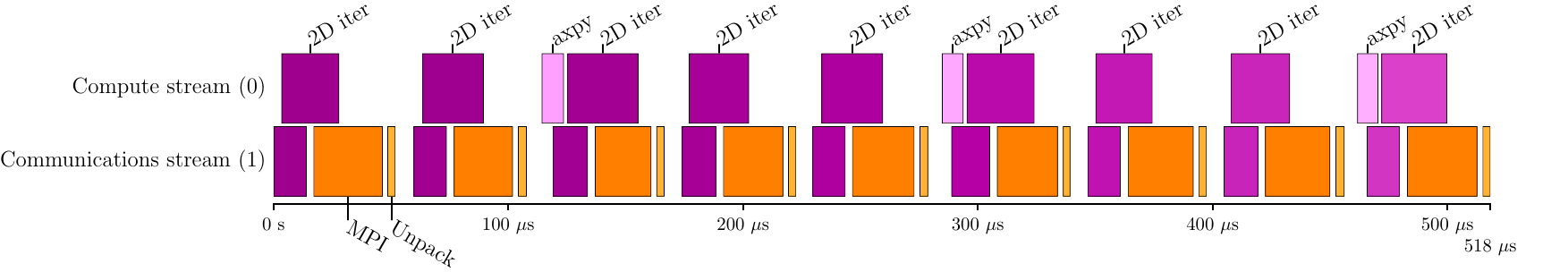}
    \caption{Timeline of both the Compute stream and Communications stream for 3 iterations of the 2D external mode. Contrary to the 3D kernels shown in Figure~\ref{fig:overview_u_mpi}, the packing is done by the computation kernel and not separately.}
    \label{fig:overview_u_2d_mpi}
\end{figure}

Kernels used for the 2D mode execute in much shorter time, and thus the latency of the MPI communications starts to become a bottleneck. Figure~\ref{fig:overview_u_2d_mpi} shows a typical timeline for a few iterations of the 2D mode where the white gaps between the colored blocks indicate an idling GPU. 
Since communications now take a significant amount of time, it is often no longer possible to launch the next kernels in advance, and the kernel launch latency cannot be hidden. To partially mitigate this problem, the boundary computation kernel is also responsible for the packing in the MPI buffers, which saves a few microseconds of latency.

This duality in MPI regimes, with heavy compute kernels for the 3D components, and much lighter 2D kernels where the communication cost is significant, will be the determinant factor in scaling the code to multiple GPUs. At the limit, when the strong scaling starts to weaken, the 2D mode takes an almost constant amount of time, due to the latency of everything, while the 3D continues to scale. This is also whey our timings are a very good fit for Amdahl's law.

\clearpage
\section{Performance and scaling of the model}\label{sec:perf_results}
On CPUs, explicit models typically exhibit predictable performance, with iteration time scaling approximately linearly with problem size. In contrast, GPU performance depends strongly on workload characteristics due to the massively parallel architecture. In particular, small problem sizes lead to underutilization of the device. This effect is further exacerbated in multi-GPU configurations with MPI, even before accounting for communication overheads. This sensitivity to workload size motivates a detailed examination of single- and multi-GPU performance.

\subsection{Single-GPU performance}
We first examine single-device performance across architectures, focusing on both scaling behavior and effective hardware utilization. A large GPU--CPU speedup alone does not necessarily indicate an efficient GPU implementation, as it may reflect limitations of the CPU baseline. In this work, the same codebase is used for both CPU and GPU targets, although the optimizations are primarily GPU-oriented.

\begin{figure}[htpb]
    \centering
    \includegraphics[width=\textwidth]{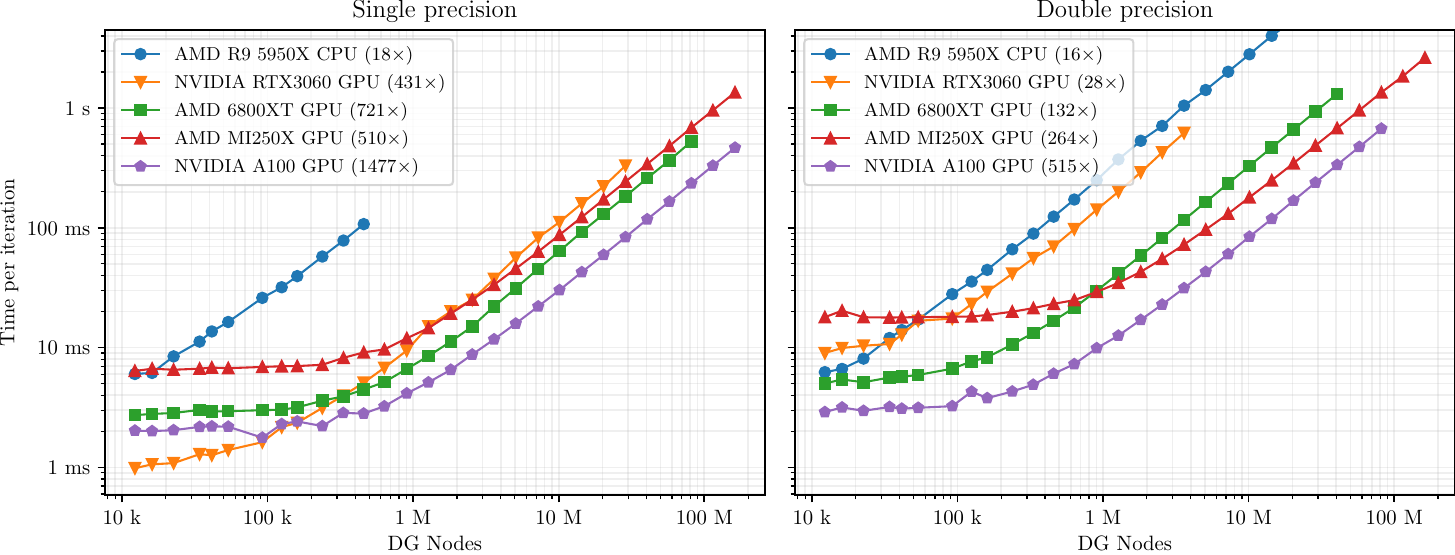}
    \caption{Performance of the 3D model on various hardware platforms with 32 layers and increasing horizontal resolution. The multiplier reported in the legend denotes the speedup relative to a hypothetical single-core CPU operating in double precision for a sufficiently large problem. In practice, the 16-thread execution on the AMD R9 5950X is used as the reference and assigned a multiplier of $16\times$.}
    \label{fig:scaling_global_alldev}
\end{figure}

A comparison of single-device performance across several architectures is shown in Figure~\ref{fig:scaling_global_alldev}. The figure includes results for a 16-core AMD R9 5950X CPU (using 16 threads), consumer GPUs from NVIDIA (RTX~3060) and AMD (RX~6800XT), and data-center GPUs from NVIDIA (A100) and AMD (MI250X). The MI250X consists of two Graphics Compute Dies (GCDs), each exposed as a separate device at the programming level; results reported here correspond to a single GCD.

While double-precision performance is consistently lower than single precision, three distinct categories emerge. On CPUs, the performance degradation is modest, as expected for a code dominated by scalar arithmetic, for which the cost of single and double precision is similar. Data-center GPUs typically exhibit FP32/FP64 throughput ratios of $1/2$ (e.g., A100) or even $1$ (MI250X), resulting in double-precision performance between one-third and one-half of FP32. Memory bandwidth also plays a significant role, which explains why the MI250X is approximately twice as slow in double precision despite having identical peak throughput in FP32 and FP64. 

In contrast, consumer GPUs have FP32/FP64 ratios ranging from $1/16$ to $1/64$, resulting in a pronounced degradation in double-precision performance. Nevertheless, GPUs provide substantial overall performance gains: a laptop RTX~3060 can achieve performance comparable to approximately 430 CPU cores, an MI250X GCD exceeds 500 cores, and an A100 approaches 1500 cores.

The resulting implementation makes effective use of hardware resources. Memory-bound kernels sustain up to 80\% of peak bandwidth, while compute-bound kernels reach approximately 60\% of peak floating-point throughput. Many kernels exhibit mixed behavior due to kernel fusion, combining memory-bound and compute-intensive operations.

\begin{figure}[htpb]
    \centering
    \includegraphics[width=\textwidth]{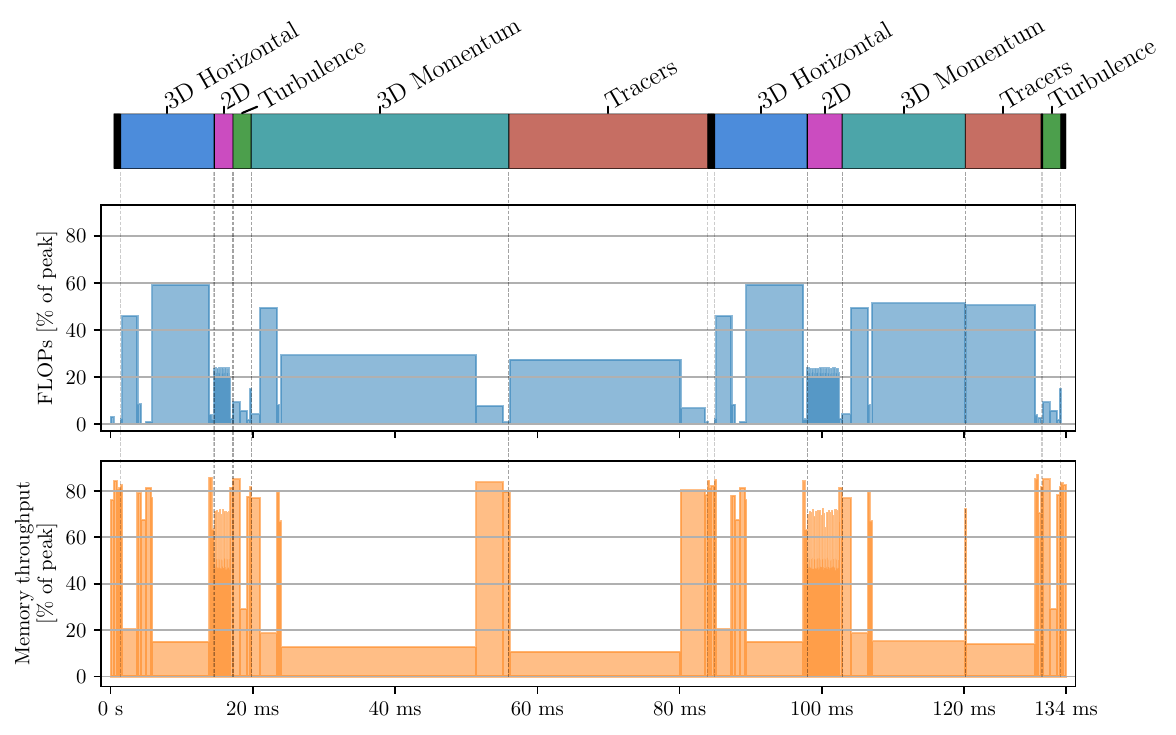}
    \caption{Global memory bandwidth and floating-point throughput as a percent of peak over a complete time-step. The sustained average is about 30\% for both. Measurements performed on an A100 GPU using single precision.}
    \label{fig:percent_peak}
\end{figure}
Figure~\ref{fig:percent_peak} shows the compute and bandwidth utilization during a typical time step. Sustained averages of approximately 30\% of peak for both compute and bandwidth are notable for a low-order method, and reflect a high degree of efficiency despite the combination of memory-bound and compute-intensive kernels.

\subsubsection*{Non-linear scaling with horizontal resolution}
While the code exhibits near-linear scaling at large mesh sizes, reducing the horizontal resolution does not necessarily lead to proportional performance gains. As shown in Figure~\ref{fig:scaling_global_alldev}, the iteration time on GPUs becomes nearly constant below a threshold between $10^5$ and $10^6$ nodes (corresponding to approximately 500 to 5000 triangles with 32 layers).

A 3D step involves many 2D computations, which are significantly faster than the 3D kernels. Consequently, the 2D component is the first to underutilize the GPU as the resolution decreases. In this regime, the 2D execution time is dominated by kernel launch latency and therefore remains effectively constant. Since the 3D kernels are substantially more expensive, their latency overhead is negligible in comparison.

\subsubsection*{Non-linear scaling with vertical resolution}
\begin{figure}[htpb]
    \centering
    \includegraphics[width=\textwidth]{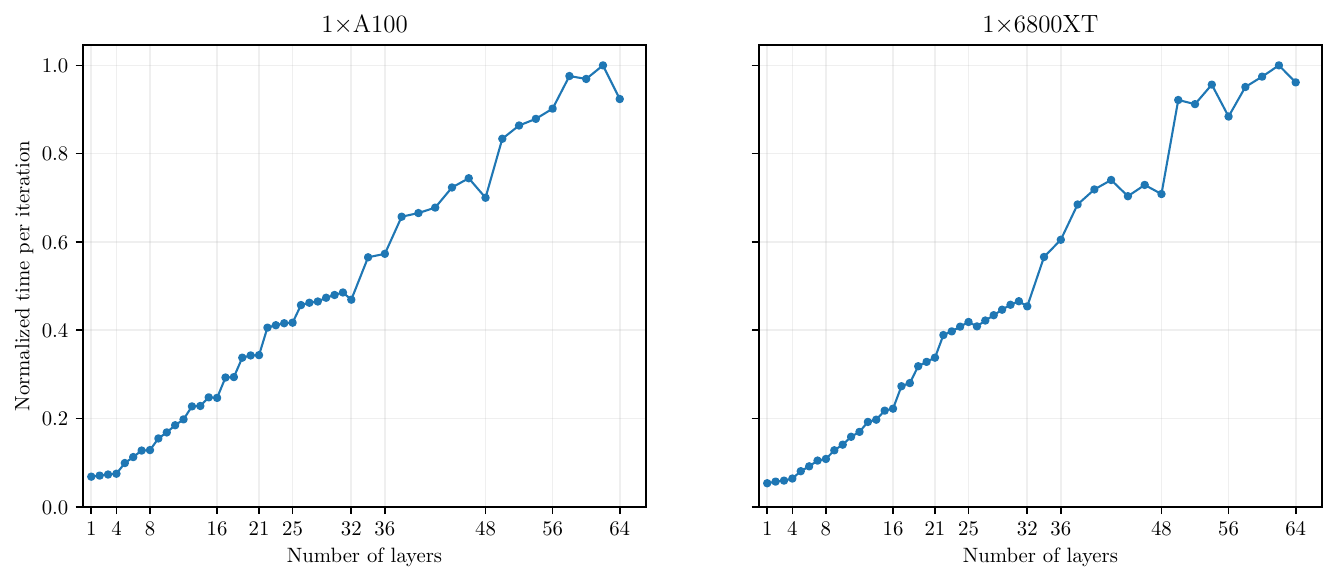}
    \caption{Normalized time per iteration of the 3D scheme as a function of the number of layers for an NVIDIA A100 GPU (left) and AMD 6800XT GPU (right) using single precision. The visible dips at values like 16, 32 and 64 are numbers that allow all threads from a block to be used, and also have a good coalescence in the memory accesses.}
    \label{fig:time_layer_A100-6800XT}
\end{figure}

Due to the block-structured execution, the time per step does not scale linearly with the number of layers, as shown in Figure~\ref{fig:time_layer_A100-6800XT}. As described in Section~\ref{sec:thread_assignment}, each thread block processes all layers of its assigned columns. This can lead to partial thread utilization. For instance, with 20 layers, an efficient configuration assigns 6 columns per block, yielding 120 active threads out of 128. Moreover, memory writes occur in chunks equal to the number of columns, here 6, which is suboptimal for memory coalescence. For larger layer counts, columns are split vertically and processed sequentially by threads within the block from top to bottom. As one can expect, a number of layers that perfectly divides the block size will often result in better performance.

Because each block processes all layers of its assigned columns, sharing 2D data among threads within a column is advantageous. This is implemented using shared memory. To reduce complexity and maximize performance, only statically allocated shared memory is used, with its size fixed at compile time. Consequently, an upper bound on the number of columns per block must be specified in advance, and shared memory for 2D data is allocated accordingly. In SLIM, this maximum is set to 32 columns per block, with a block size of 128 threads. As a result, when fewer than 4 layers are present, the total number of elements per block is below 128, leaving some threads idle. This accounts for the nearly constant runtime observed for 1 to 4 layers in Figure~\ref{fig:time_layer_A100-6800XT}.

\clearpage
\subsection{Multi-GPU scaling}
SLIM is designed for distributed-memory execution using MPI, with one GPU per rank. The scaling behavior closely mirrors that observed on a single GPU, with latency-dominated components limiting performance at small workloads. On a single GPU, this latency is primarily due to kernel launches; in the multi-GPU setting, it additionally includes MPI communication overheads and the cost of packing and unpacking data. This overhead is partially mitigated through the overlap strategy described in Section~\ref{sec:multi_gpu}. However, due to the relatively low computational cost of the 2D component, its associated latency cannot be fully hidden and remains the dominant limitation for strong scaling.

As a result, the overall scaling of the 3D model is well described by Amdahl's law. The latency associated with the 2D component effectively acts as the sequential fraction of the computation, while the 3D component exhibits near-ideal strong scaling.

Benchmarks were conducted on two EuroHPC systems. The first is MeluXina, equipped with 4 NVIDIA A100 (40\,GB) GPUs per node, using a non GPU-aware OpenMPI implementation. Within each node, GPUs are interconnected via NVLink, while communication with the host and network relies on PCIe. 

The second system is LUMI (GPU partition LUMI-G), equipped with 4 AMD MI250X GPUs per node, each providing 128\,GB of memory. Each MI250X consists of two Graphics Compute Dies (GCDs), exposed as separate devices at the programming level; results are therefore reported in terms of the number of GCDs. Within a node, MI250X modules are connected through high-speed GPU--GPU links, and each GPU is directly connected to the Slingshot-11 interconnect, providing 200\,Gb/s per network endpoint. GPU-aware Cray MPICH is used on this system.

\begin{figure}[htpb]
    \centering
    \includegraphics[width=\textwidth]{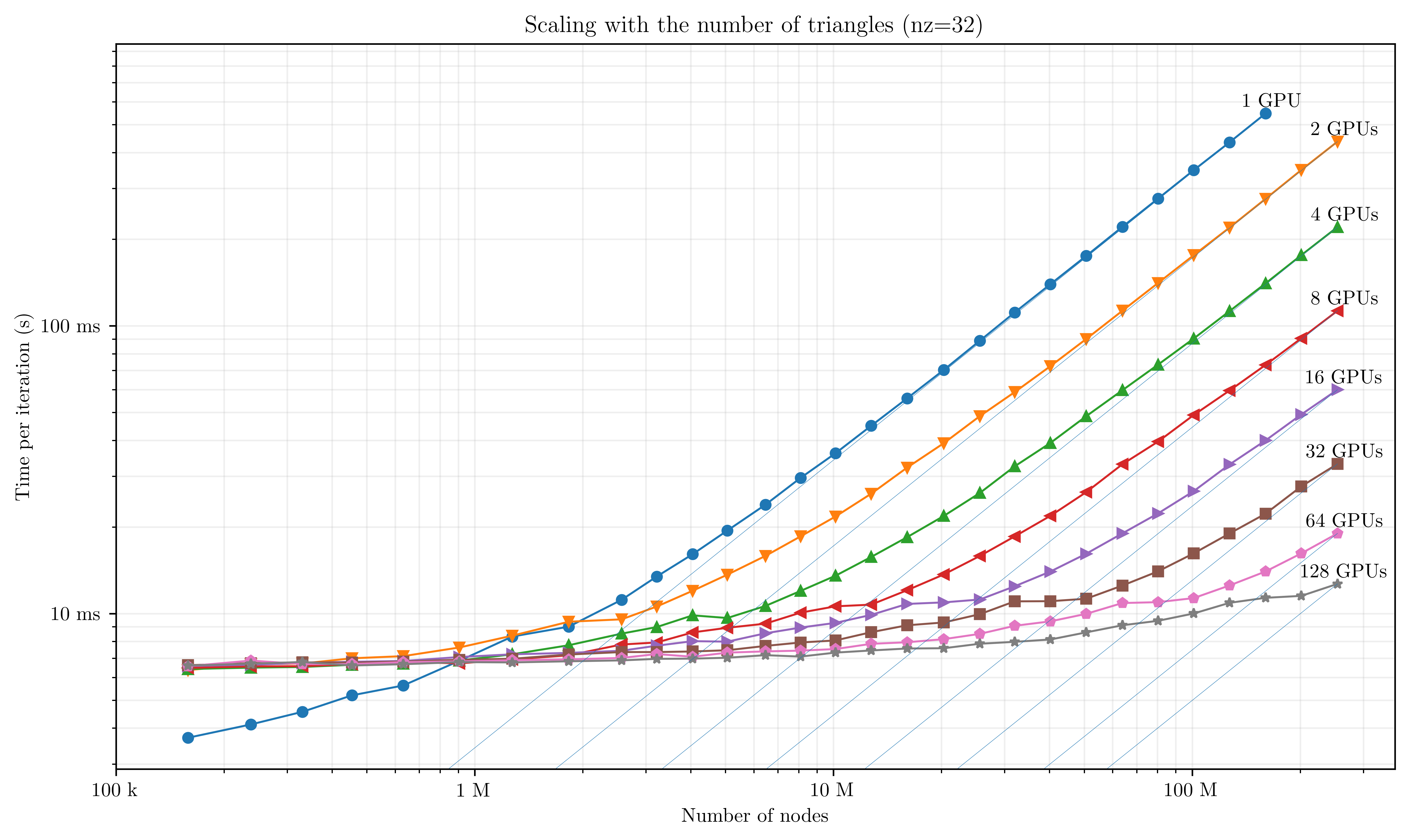}
    \caption{Scaling of the 3D model with 32 layers on the MeluXina cluster with A100 GPUs.}
    \label{fig:scaling_global_A100_32l}
\end{figure}

Figure~\ref{fig:scaling_global_A100_32l} illustrates the scaling on the MeluXina cluster with A100 GPUs. Two distinct latency regimes can be observed: on a single GPU, the minimum time per iteration is approximately \SI{3}{\milli\second}, while it increases to about \SI{6}{\milli\second} as soon as multiple GPUs are used. Importantly, this latency remains essentially constant as the number of GPUs increases, indicating that network contention is not a limiting factor.

This configuration uses a time step ratio of 20 between the internal and external modes, i.e., one internal iteration corresponds to 20 external iterations, which is representative of typical applications. Each internal iteration involves approximately 100 halo exchanges, about 90\% of which originate from the 2D mode. This corresponds to roughly 100 stream synchronizations, 100 MPI send/receive operations, and 200 additional kernel launches. Assuming comparable costs for these operations, the additional \SI{3}{\milli\second} per iteration corresponds to an average overhead of approximately \SI{7.5}{\micro\second} per synchronization, communication, or kernel launch, consistent with the expected order of magnitude.

\begin{figure}[htpb]
    \centering
    \includegraphics[width=\textwidth]{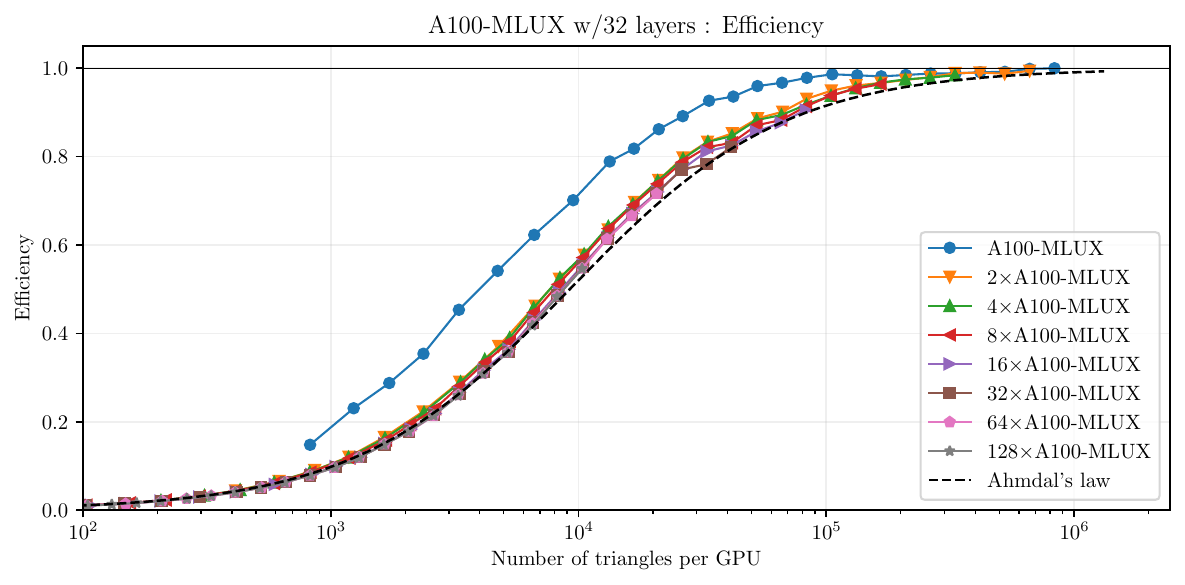}
    \caption{Efficiency of the 3D model with 32 layers on the MeluXina cluster with A100 GPUs.}
    \label{fig:efficiency_A100-MLUX-32}
\end{figure}

Figure~\ref{fig:efficiency_A100-MLUX-32} shows the efficiency as a function of the number of elements per GPU. For two or more GPUs, the efficiency depends primarily on this quantity, indicating near-ideal weak scaling. The observed behavior is also well described by Amdahl's law, as expected.

\begin{figure}[htpb]
    \centering
    \includegraphics[width=\textwidth]{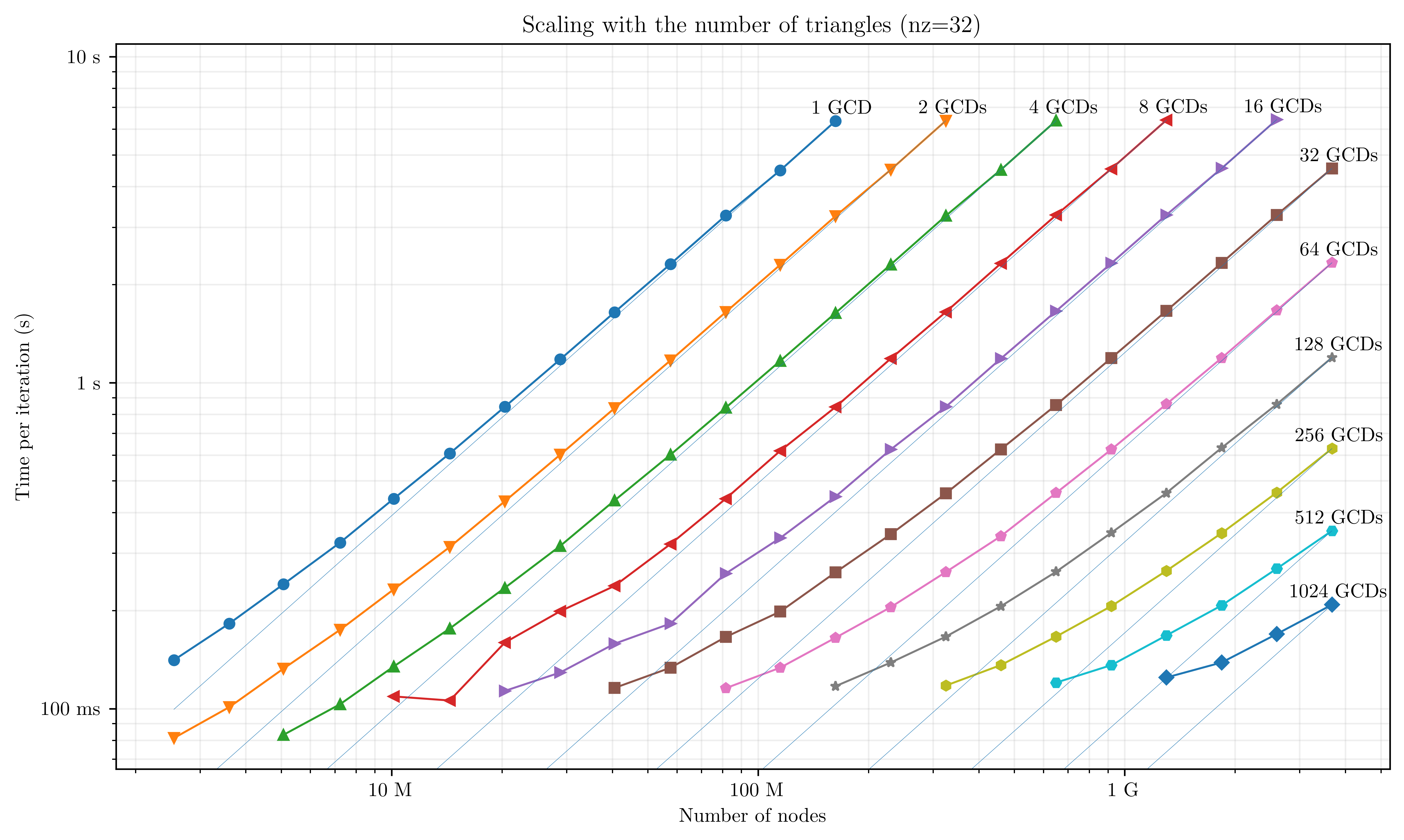}
    \caption{Scaling of the 3D model with 32 layers on the LUMI cluster equipped with MI250X GPUs.}
    \label{fig:scaling_global_MI250X_32l}
\end{figure}

Figure~\ref{fig:scaling_global_MI250X_32l} illustrates the scaling on the LUMI cluster with MI250X GPUs. Owing to the larger scale of the system, experiments were conducted with significantly larger configurations, reaching up to 1024 Graphics Compute Dies (GCDs), corresponding to 512 GPUs across 128 nodes. Although efficiency decreases at large scale, the model continues to scale provided the problem size per GPU remains sufficiently large.

As in many parallel applications, SLIM involves a trade-off between time to solution and resource efficiency. An efficiency of approximately 80\% is often considered a practical target. For SLIM, this corresponds to about $4\times 10^4$ triangles per GPU, or approximately $5\times 10^6$ nodes globally. Given that the intended use case involves simulations with at most a few million triangles distributed over 32 to 64 GPUs, the current implementation achieves satisfactory performance.

\clearpage
\section{Application to the Great Barrier Reef}\label{sec:applications}
To demonstrate that the model is applicable beyond idealized benchmarks and can handle realistic coastal domains, we consider a high-resolution simulation of the Great Barrier Reef (GBR).

The circulation in the GBR is driven by a combination of large-scale currents, winds, tides, and complex bathymetry. In particular, the South Equatorial Current (SEC) bifurcates near 14--18$^\circ$S into the northward Coral Sea Coastal Current and the southward East Australian Current \cite{lambrechtsMultiscaleModelHydrodynamics2008}. These currents, modulated by local forcing, control reef flushing, larval connectivity, and the transport of nutrients and pollutants. In addition, small-scale features such as tidal jets and eddies (from \SI{100}{\meter} to a few kilometers) play a key role in reef-scale processes.

Modeling these interacting processes across a wide range of spatial scales remains challenging. Uniform-resolution models in the GBR, typically ranging from \SI{1.5}{\kilo\meter} to \SI{4}{\kilo\meter}, are unable to resolve these fine-scale dynamics. The present approach addresses this limitation using a highly non-uniform mesh, refined near reefs and progressively coarsened offshore, enabling the representation of both basin-scale circulation and reef-scale processes.

\begin{figure}[htpb]
    \centering
    \includegraphics[width=\textwidth]{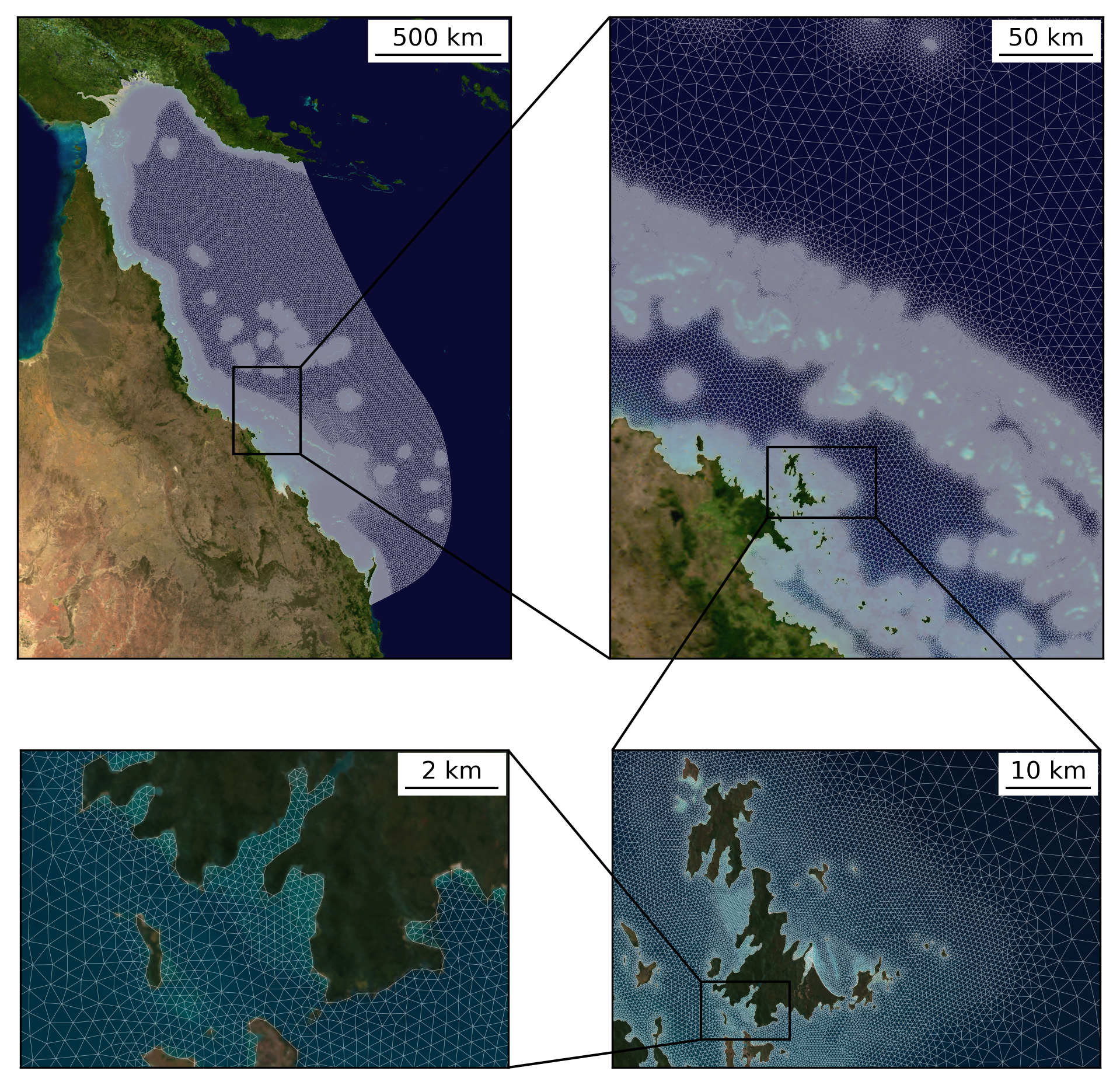}
    \caption{Computational mesh for the Great Barrier Reef configuration. The horizontal resolution varies from \SI{200}{\meter} in reef regions to \SI{10}{\kilo\meter} in the open ocean, enabling the resolution of fine-scale coastal dynamics while maintaining tractable computational cost offshore. Insets highlight the progressive refinement near complex topographical structures.}
    \label{fig:mesh_zoom_gbr}
\end{figure}

Figure~\ref{fig:mesh_zoom_gbr} shows the computational domain and horizontal mesh. The mesh consists of 3.3 million triangles, with resolution ranging from \SI{200}{\meter} near reefs to \SI{10}{\kilo\meter} offshore. The 2D mesh is extruded vertically into a prismatic grid with 10 layers in shallow regions and up to 29 layers in deeper areas, resulting in approximately 34 million elements. The resulting resolution enables the model to capture a wide range of flow features, as illustrated below.

\begin{figure}[htpb]
    \centering
    \includegraphics[width=\textwidth]{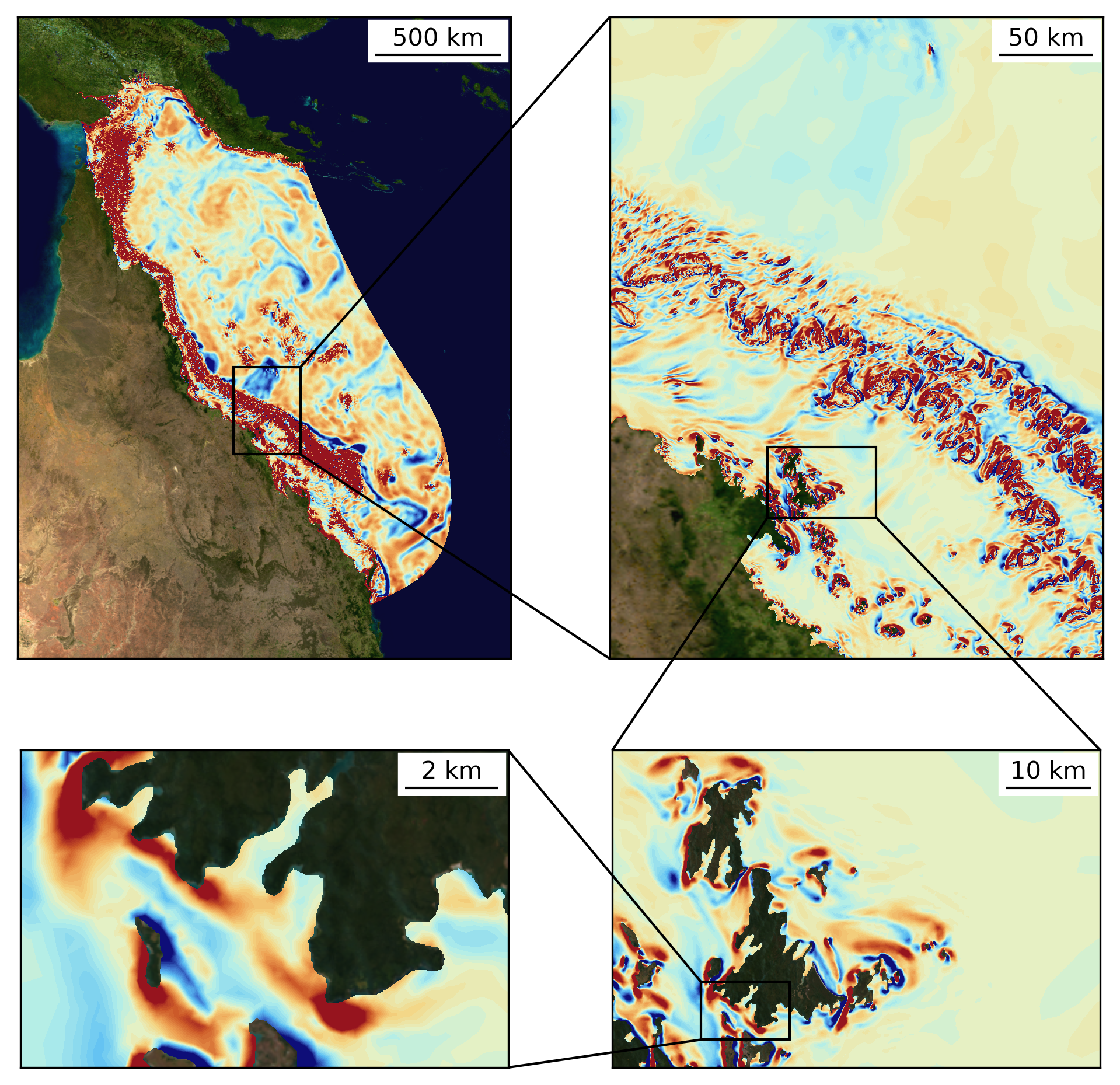}
    \caption{Surface vertical vorticity at increasing levels of zoom in the Great Barrier Reef domain. The solution reveals fine-scale flow structures down to scales of a few hundred meters, illustrating the ability of the model and mesh to capture small-scale dynamics induced by the topography.}
    \label{fig:surf_vort_gbr}
\end{figure}

\begin{figure}[htpb]
    \centering
    \includegraphics[width=\textwidth]{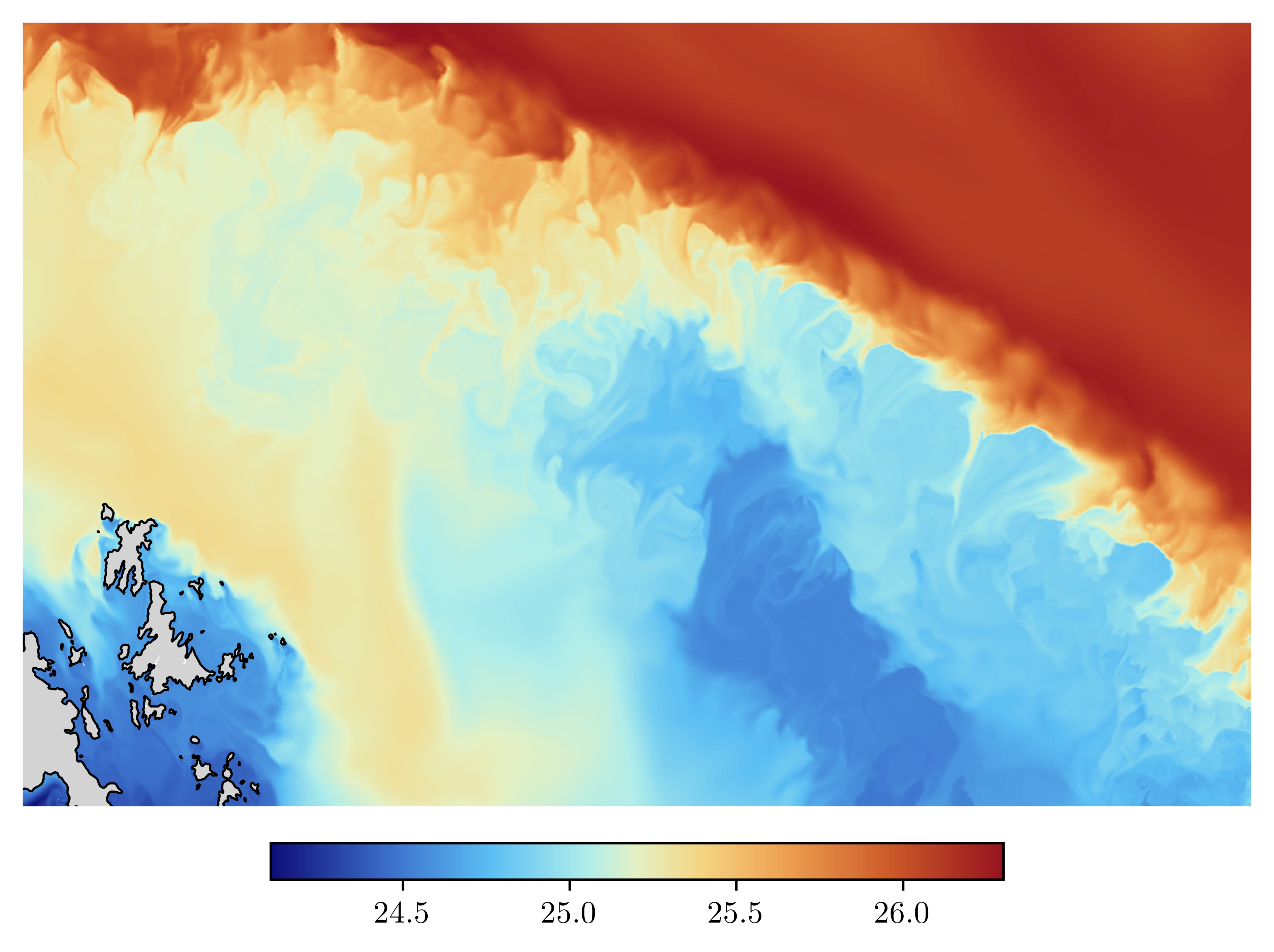}
    \caption{Modelled sea surface temperature on October 31, 2024 at 10:00, after two months of simulation. The solution resolves fine-scale spatial variability associated with reef topography, including signatures of tidal jets and localized mixing processes.}
    \label{fig:surf_temp_gbr}
\end{figure}

Figure~\ref{fig:surf_vort_gbr} shows the surface vertical vorticity at increasing levels of zoom, highlighting flow structures down to scales of a few hundred meters. Mesh resolution has a strong impact on coral connectivity, as demonstrated by \cite{saint-amandBiophysicalModelsResolution2023}. Figure~\ref{fig:surf_temp_gbr} illustrates the fine-scale variability in the surface temperature field, capturing features associated with complex reef topography. In particular, resolving tidal jets and their influence on local flow patterns requires sufficiently fine spatial resolution.

The simulation was performed on 32 AMD MI250X GPUs (64 GCDs) in double precision, achieving a throughput of approximately 100 simulated days per day of wall-clock time. This demonstrates that high-resolution coastal simulations of this scale are computationally feasible on modern GPU systems. While the present configuration is not yet calibrated for scientific analysis, it provides a representative test of model performance and scalability in a realistic setting.

The model setup relies on multiple data sources. Bathymetry is derived from 30\,m datasets for the Great Barrier Reef and Torres Strait \cite{beamanAusBathyTopoGreatBarrier2020,beamanTorresStraitBathymetry2023}, complemented by 100\,m datasets for the Coral Sea \cite{beamanHighResolutionDepth2020} and the Gulf of Papua \cite{daniellGulfPapuaBathymetry2020}. Coastlines are obtained from OpenStreetMap \cite{openstreetmapcontributorsPlanetDumpRetrieved2015}.

Open boundary conditions (temperature, salinity, and currents) are derived from the BRAN2023 reanalysis dataset \cite{chamberlainBluelinkOceanReanalysis2024}. Tidal forcing is provided by TPXO10v2 \cite{egbertEfficientInverseModeling2002}, while atmospheric forcing (wind, precipitation, and heat fluxes) is obtained from the BARRA2-C2 regional atmospheric reanalysis dataset \cite{BARRA2}, described by \cite{suBARRA2DevelopmentNextgeneration2022,suBARRAC2Development2024}. Reef extent is based on \cite{AllenCoralAtlas2020,GlobalDistributionCoral2021}. The density $\rho$ is computed from the equation of state $\rho(S,T,p)$ following \cite{jackettAlgorithmsDensityPotential2006}.

\clearpage
\section{Conclusion}\label{sec:conclusion}
We demonstrated that unstructured-mesh ocean models based on the Discontinuous Galerkin finite element method can effectively leverage modern GPU architectures. The DG-FE formulation is inherently well suited to GPU execution, enabling hardware utilization levels that can rival structured-grid models. The resulting multi-GPU implementation scales efficiently from consumer laptops to large HPC clusters with hundreds of devices. This performance gain  makes it feasible to perform three-dimensional coastal simulations at resolutions and scales that were previously difficult to attain.

The DG-FE formulation offers properties that align naturally with GPU architectures: high data locality, element-wise independence in explicit computations, and a high arithmetic intensity relative to other unstructured-mesh approaches. These properties allowed us to sustain up to 80\% of peak memory bandwidth for memory-bound kernels and approximately 60\% of peak floating-point throughput for compute-bound kernels, with a sustained average of about 30\% of peak compute and memory throughput over a complete time step. These figures are notable for a low-order method on an unstructured mesh, and are comparable to utilization rates reported for structured-grid GPU models. For example, Veros achieves strong performance gains through JAX just-in-time compilation \cite{hafnerFastCheapTurbulent2021} but on a regular grid, while Oceananigans.jl reaches high efficiency through kernel-fusion on a structured finite-difference grid \cite{silvestriGPUBasedOceanDynamical2025}. Our results demonstrate that comparable efficiency can be achieved on unstructured meshes when the DG formulation is exploited through appropriate data layouts (structure-of-arrays with Hilbert curve reordering), matrix-free solvers for vertically structured operators, and a dedicated cell layout for column-wise implicit solves.

Our implementation achieved efficient scaling from a single consumer GPU to 1024 HPC-grade devices across two state-of-the-art EuroHPC systems (MeluXina and LUMI). A single NVIDIA A100 delivers performance equivalent to approximately 1500 CPU cores, and replacing a 128-core CPU node with a 4×A100 GPU node yields a speedup of approximately 50. These gains are preserved as the number of devices increases: weak-scaling efficiency remains high, and strong scaling follows Amdahl's law closely, with the latency-dominated 2D external mode acting as the effectively sequential fraction. This performance is obtained from a single codebase targeting CPU, CUDA, and HIP backends through a lightweight abstraction layer, ensuring portability across architectures with minimal overhead.
This scaling behavior reflects the dual nature of the mode-splitting approach common to most ocean models \cite{shchepetkinRegionalOceanicModeling2005, karnaThetisCoastalOcean2018, madecNEMOOceanEngine2022}. The 3D baroclinic component, which dominates the computational cost, exhibits near-ideal scaling owing to its high arithmetic intensity and effective computation-communication overlap. In contrast, the 2D barotropic mode, while computationally cheap, involves many short kernels and frequent halo exchanges, making it sensitive to MPI latency and kernel launch overhead. This dichotomy has also been observed in GPU implementations of structured-grid models \cite{xuGpuPOMGPUbasedPrinceton2014}. Similar challenges are reported with the barotropic mode in LICOM3-HIP \cite{weiAcceleratingLASGIAP2024}. The overlap strategy employed here, where boundary elements are processed first to initiate asynchronous halo exchanges while interior computations proceed, partially mitigates this issue but cannot fully hide the latency at extreme strong-scaling limits.

GPU acceleration makes high-resolution, three-dimensional coastal simulations over large domains computationally feasible. The Great Barrier Reef application, with mesh resolution down to 200 m near the reefs and 34 million prisms, ran on 32 MI250X GPUs (64 GCDs) while maintaining a physical-to-numerical time ratio of approximately 100. This represents the first application of a 3D model to the entire GBR with a sub-reef scale resolution. Such a model can resolve fine-scale features such as tidal jets and reef-scale eddies that are critical for larval connectivity, nutrient transport, and sediment dynamics.

This capability addresses a longstanding bottleneck in coastal ocean modeling. While unstructured-mesh models have long been recognized for their geometrical flexibility, their higher computational cost per degree of freedom compared to structured-grid models has historically limited their practical application \cite{danilovFinitevolumESeaIce2017}. The GPU acceleration presented here effectively removes this constraint for regional and coastal domains. The present implementation brings the cost of high-resolution unstructured simulations in line with what is routinely achieved on structured grids at coarser resolution, potentially opening the door to operational applications in environmental management.

Some limitations should nonetheless be acknowledged. The performance scaling with mesh resolution is not strictly linear. GPU utilization saturates below approximately $10^6$ DG nodes, corresponding to roughly 5000 triangles with 32 layers, below which the device is underutilized and iteration time becomes dominated by kernel launch latency. Similarly, performance exhibits non-monotonic behavior with the number of vertical layers due to thread-block occupancy constraints: layer counts that do not evenly divide the block size result in idle threads and suboptimal memory coalescence. While these effects are well understood and primarily affect small simulations, they imply that the model is most efficient for medium-to-large problems with layer counts aligned to powers of two. Furthermore, the Great Barrier Reef simulation, while demonstrating the model's scalability and resolution capabilities, has not been calibrated or validated against observational data. The results should therefore be interpreted as a computational feasibility demonstration rather than a scientifically validated study of GBR dynamics.

More broadly, this work shows that the DG-FE approach, long considered too expensive for routine use in ocean modeling, can become computationally competitive with structured-grid methods when GPU architectures are effectively exploited. As exascale computing systems become more widely available and GPU architectures continue to evolve, the inherent advantages of unstructured meshes, geometrical flexibility, local refinement, and multi-scale resolution, may increasingly be leveraged without the traditional computational overhead. In this context, GPU-accelerated DG models emerge as a promising tool for the next generation of coastal and regional ocean simulations, where high resolution, complex topography, and multi-physics coupling are simultaneously required.

\clearpage
\section*{Open Research}

The SLIM model source code used in this study is openly developed at \url{https://git.immc.ucl.ac.be/slim/slim4} and is distributed under the GNU General Public License v3.0 or later (GPLv3+). Documentation, installation instructions, and examples are available at \url{https://slim.git-page.immc.ucl.ac.be/slim4/v-0.9/index.html}, while an overview of the model and publications using SLIM are provided at \url{https://www.slim-ocean.be/}. A permanent, citable archive of the version of SLIM used for this manuscript, together with the scripts and input files needed to reproduce the numerical experiments and figures, will be deposited in a public repository such as Zenodo and assigned a DOI before publication.

The synthetic benchmark configurations used to evaluate the multi-GPU implementation are available in the SLIM benchmark example at \url{https://slim.git-page.immc.ucl.ac.be/slim4/v-0.9/examples/Benchmark.html}. These benchmarks are designed to assess computational performance and do not rely on external geophysical data.

The Great Barrier Reef (GBR) model configuration used for the realistic performance test is available in the SLIM example repository at \url{https://slim.git-page.immc.ucl.ac.be/slim4/v-0.9/examples/Great_Barrier_Reef_testcase.html}. This configuration is provided to reproduce the performance-oriented test case reported in this manuscript. It was not calibrated or validated for scientific analysis of GBR hydrodynamics, and should therefore not be interpreted as a research-grade GBR circulation simulation.

The public GBR example uses bathymetry, coastline, atmospheric forcing, ocean boundary conditions, and coral reef distribution data obtained from public data providers. The bathymetric data are from the Geoscience Australia Great Barrier Reef and Coral Sea depth products \cite{beamanAusBathyTopoGreatBarrier2020, beamanHighResolutionDepth2020, beamanTorresStraitBathymetry2023} and the Gulf of Papua bathymetry data set \cite{daniellGulfPapuaBathymetry2020}. Coral reef distribution data are from the Allen Coral Atlas \cite{AllenCoralAtlas2020} and the global coral reef distribution data set compiled by UNEP-WCMC, WorldFish Centre, WRI, and TNC \cite{GlobalDistributionCoral2021}. Coastline data are from OpenStreetMap \cite{openstreetmapcontributorsPlanetDumpRetrieved2015}. Atmospheric forcing is from BARRA2, the Australian Regional Atmospheric Reanalysis \cite{BARRA2}, and ocean boundary and initial conditions are from BRAN2023, the Bluelink Ocean Reanalysis \cite{chamberlainBluelinkOceanReanalysis2024}. Access conditions and licenses are those specified by the respective data providers.

The original TPXO tidal forcing data used in the internal GBR performance tests cannot be redistributed, because access to TPXO products must be requested directly from the data provider. The archived and public GBR example therefore uses simplified tidal forcing with realistic magnitudes. This substitution does not affect the performance conclusions of the manuscript, because the GBR case is used only to benchmark the implementation under realistic mesh, bathymetry, and forcing complexity, not to draw scientific conclusions about tides or circulation in the Great Barrier Reef.

\subsection*{Acknowledgments}
This work was supported by the Special Research Fund (FSR) - UCLouvain.

Computational resources have been provided by the supercomputing facilities of the Université catholique de Louvain (CISM/UCL) and the Consortium des Équipements de Calcul Intensif en Fédération Wallonie Bruxelles (CÉCI) funded by the Fond de la Recherche Scientifique de Belgique (F.R.S.-FNRS) under convention 2.5020.11 and by the Walloon Region.

The present research also benefited from computational resources made available on Lucia, the Tier-1 supercomputer of the Walloon Region, infrastructure funded by the Walloon Region under the grant agreement n°1910247.

The simulations on multiple NVIDIA GPUS were performed on the Luxembourg national supercomputer MeluXina. The authors gratefully acknowledge the LuxProvide teams for their expert support.

The authors acknowledge LUMI-BE for awarding this project access to the LUMI supercomputer, owned by the EuroHPC Joint Undertaking, hosted by CSC (Finland) and the LUMI consortium through a LUMI-BE Regular Access call. LUMI-BE is joint effort from BELSPO (federal), SPW Économie, Emploi, Recherche (Wallonia), Department of Economy, Science \& Innovation (Flanders) and Innoviris (Brussels).

The authors declare there are no conflicts of interest for this manuscript.

\printbibliography

\end{document}

\clearpage

\maketitle

\vspace{0.5cm}
\section*{Spatial Discretization}\label{sec:appendix}
The spatial discretization used in this article is based on the Discontinuous Galerkin (DG) method with linear nodal basis functions. Since the 3D mesh is constructed by vertically extruding the 2D mesh, the corresponding basis functions $\phi$ are defined as the tensor product of the 2D horizontal basis functions $\phih$ and a vertical 1D basis function $\phiz$:
\[
    \phi(\xi, \eta, \zeta) = \phih(\xi, \eta) \cdot \phiz(\zeta) \; .
\]
When using nodal basis functions, we sometimes refer to a single 3D index $i$, or to a pair of indices $(i\idxh, i\idxz)$ corresponding to the horizontal and vertical indices, respectively:
\[
    \phi^i \defeq \phi^{i\idxh, i\idxz}(\xi, \eta, \zeta) \defeq \phih^{i\idxh}(\xi, \eta)\, \phiz^{i\idxz}(\zeta) \; .
\]

Since the top and bottom boundaries of the 3D elements may not be horizontal, the vertical coordinate $\zeta$ is a function of the physical coordinates $(x, y, z)$, while $\xi$ and $\eta$ do not depend on $z$.

Using the DG method introduces discontinuities (jumps) in discretized fields at element interfaces. We therefore define the following operators for the mean value, mean difference, maximum, and upwind value on an interface:
\begin{align*}
    \bmean{\alpha} &= \dfrac{\alpha\inte + \alpha\exte}{2}, 
    & \quad  
    \bdiff{\alpha} &= \dfrac{\alpha\inte - \alpha\exte}{2}, \\[2mm]
    \bmax{\alpha}  &= \max(\alpha\inte, \alpha\exte), 
    & \quad  
    \alpha\upw &= 
    \begin{cases*}
        \alpha\inte, & if $\nn \cdot \uuu^* \ge 0$, \\
        \alpha\exte, & if $\nn \cdot \uuu^* < 0$,
    \end{cases*}
\end{align*}
for any field $\alpha$, where $\alpha\inte$ and $\alpha\exte$ denote the interior and exterior values, respectively. The vector $\uuu^*$ is the discrete advection velocity at the interface, and $\nn$ is the unit outward normal vector. When a quantity is taken in its upwind form, it is always multiplied by the velocity field from which it originates.

Since SLIM uses triangular meshes in 2D, the 3D elements are prisms, denoted by $\prism$, with the corresponding triangle denoted $\tri$. The top and bottom faces of a prism are referred to as $\gtop$ and $\gbot$, collectively called the horizontal boundaries $\gtb = \gtop \cup \gbot$. The lateral boundaries are denoted $\glat$, and the full prism boundary is $\gfull$. In 2D, the edges of the triangle $\tri$ are denoted $\edge$. When integrating over the parent element, a ``hat'' is used to distinguish it: $\prismh$ denotes the parent prism, while $\gtbh$ and $\glath$ refer to its horizontal and lateral faces, respectively.

Because the free-surface elevation is discretized discontinuously and defines the shape of the 3D mesh, the mesh is slightly non-conformal along lateral faces. To avoid ambiguity in the integration domain, all integrals are expressed in the parent element. This change of variables introduces a jacobian, denoted as $J$ when integrating inside an element, or $\jbnd$ when integrating on its boundary. Depending on the context, $J$ and $\jbnd$ are sometimes expressed in a different form. Due to the tensor product formulation, the volume Jacobian $J$ is the product of the 2D Jacobian ($\jj$) and the vertical 1D Jacobian ($\jz$). The same is true for the lateral jacobian where $\jlat = \jz\jedge$, with $\jedge$ the jacobian of the edge of the triangle. On the top and bottom faces of the prisms, the jacobian can also be split as $\jtb = \jj/\abs{\nz}$.

Finally, to simplify notation, we denote volume integrals over a domain $\Omega$ with single angled brackets and boundary integrals over $\partial\Omega$ with double brackets:
\begin{align*}    
    \int_{\Omega} f \, \dd{\Omega} = \eint{f},
    \qquad
    \int_{\partial\Omega} f \, \dd{\sigma(\Omega)} = \eeint{f}.
\end{align*}

% Scalar variables will be written in regular math fonts while vector components will use bold fonts. When writing equations involving the gradient of a vector field, the resulting tensor will be understood as
%  \[
%     [\nabla \uuu]_{ij} = \fpart{\uuu_j}{x_i} = \partial_i \uuu_j
%  \]
% i.e.\ the variable along which the derivative is computed is indicated by the first index of the resulting tensor. To simplify notations later, we will make a distinction between the 2D and 3D gradients. The nabla operator in Cartesian coordinates written as $\nabla \triangleq [\partial_x, \partial_y, \partial_z]$ will refer to the 3D gradient, while the horizontal gradient will be written $\gradh \triangleq [\partial_x, \partial_y]$.

\subsection{2D equations}
All discrete equations are derived from the weak form of their corresponding continuous equations, using the same set of basis functions $\phi$ for both the test and trial spaces. During the derivation, interface terms that have not yet been explicitly defined are marked with an asterisk ($^*$). 
\vspace{0.3cm}
\subsubsection{Free Surface}
We start with the free-surface equation~\eqref{eq:mass2d}, which forms the first component of the external mode:
\begin{equation}\label{eq:mass2d}
    \fparti{\eta}{t} = - \gradh \cdot \Transp + s
\end{equation}
Its weak form, after integrating by parts, is
\[
    \eint{\phi\fparti{\eta}{t}} = \eint{\gradh\phi \cdot \Transp} - \eeint{\phi \nn \cdot \Transp^*} + \eint{\phi s},
\]
where $\Transp^*$ denotes the Lax--Friedrichs numerical flux across the interface, defined as $ \Transp^* = \bmean{\Transp} + \nn\,\bmax{c}\,\bdiff{\eta}$. Expressed in the parent element, this yields the discrete form of the free-surface equation:
\begin{equation}\label{eq:mass2d_discrete}
    \begin{split}        
        \eint{\phi \jj \fparti{\eta}{t}}_{\trih} 
        &= \eint{\jj\gradh \phi \cdot \Transp}_{\trih} \\
        &- \eeint{\phi (\nn \cdot \bmean{\Transp} 
        + \bmax{c} \bdiff{\eta})\jedge}_{\edgeh} \\
        &+ \eint{\phi s\jj}_{\trih}
    \end{split}
\end{equation}

\vspace{0.3cm}
\subsubsection{Depth-averaged momentum}\label{sec:momentum:2d:discrete}
The depth-averaged momentum equation~\eqref{eq:moment2d:split} constitutes the second component of the external mode:
\begin{equation}\label{eq:moment2d:split}
    \begin{split}
        \fparti{\Transp}{t} &= -g H \gradh \eta - \dfrac{H}{\rho_0} \gradh \patm + S_3 \\
        &= -g \gradh \dfrac{H^2}{2} - g H \gradh b - \dfrac{H}{\rho_0} \gradh \patm + S_3
    \end{split}
\end{equation}
Neglecting the source terms from the 3D mode and the atmospheric pressure for now, the weak form becomes
\[
    \eint{\phi\fparti{\Transp}{t}} = \eint{g (H^2/2)\gradh \phi} - \eeint{\nn \phi g (H^2/2)^*} - \eint{\phi g H \gradh b}
\]
where $(H^2/2)^*$ is also evaluated using a Lax--Friedrichs flux:
\[
    (H^2/2)^* = \bmean{H^2/2} + \nn \cdot \bmax{c}\bdiff{\Transp} \;.
\]
This leads to the discrete equation:
\[
    \begin{split}    
        \eint{\phi\fparti{\Transp}{t}} 
        &= \eint{g (H^2/2)\gradh \phi} - \eeint{\nn \phi g \bmean{H^2/2}} \\
        &- \eint{\phi g H \gradh b} - \eeint{\bmax{c}\bdiff{\Transp}}.
    \end{split}
\]
However, this formulation involves subtracting two $\bigoh(H^2)$ terms, introducing numerical errors of order $\bigoh(H^2\varepsilon_\text{machine})$. This is excessive since the total flux is only $\bigoh(H\eta)$, and $H$ is typically several orders of magnitude larger than $\eta$. To mitigate this, we rewrite $\bmean{H^2}$ as $\bmean{H^2} = (H^2)\inte - \bdiff{H^2}$ and reverse the integration by parts, yielding
\[
    \eint{\phi\fparti{\Transp}{t}} = - \eint{g H \gradh \eta} + \eeint{\nn \phi g \bdiff{H^2/2}}  - \eeint{\bmax{c}\bdiff{\Transp}} \;.
\]
Since $\bdiff{H^2}$ represents a difference of two squares, we can express it as
\[
    \bdiff{H^2} = \dfrac{(H\inte + H\exte)(\eta\inte - \eta\exte)}{2}  = 2\bmean{H} \bdiff{\eta}
\]
which significantly reduces numerical noise, yielding an error on the order of $\bigoh(H\eta\varepsilon_\text{machine})$. This ``reverse integration by parts'' procedure will be reused for other equations, though it is most impactful here.

Reintroducing the atmospheric pressure and the 3D-mode contribution $\fdddh$ (the discrete equivalent of $S_3$, discussed later), we obtain
\[
    \begin{split}
        \eint{\phi\fparti{\Transp}{t}} 
        & = - \eint{g \phi H \gradh \eta} + \eeint{\nn \phi g \bmean{H} \bdiff{\eta}} \\
        &- \eeint{\bmax{c}\bdiff{\Transp}} - \eint{\phi H/\rho_0 \gradh \patm} + \fdddh 
        \;.
    \end{split}
\]
Finally, expressing the integrals in the parent triangle yields the final discrete 2D momentum equation:
\begin{equation}\label{eq:moment2d:discrete}
    \begin{split}
        \eint{\phi\fparti{\Transp}{t} \jj}_{\trih} =
        &- \eint{g \phi H \gradh \eta \jj}_{\trih} \\
        &+ \eeint{\nn \phi g \bmean{H} \bdiff{\eta} \jedge}_{\edgeh} \\
        &- \eeint{\bmax{c}\bdiff{\Transp} \jedge}_{\edgeh}\\
        &- \eint{\phi H/\rho_0 \gradh \patm \jj}_{\trih} + \fdddh \;.
    \end{split}
\end{equation}

When coupling the internal and external modes in the temporal scheme, the average of the 2D fluxes ($\fdd$) in \eqref{eq:moment2d:discrete} is required, together with and their corresponding transport $\Transp$. Both are used an input for the vertical part of the 3D time step. Thus, during the iterations of the 2D mode, the average transport $\bar{\Transp}$ is accumulated as 
\begin{equation}
    \bar{\Transp} = \dfrac{1}{m}\sum_{\text{2D iterations}} \Transp \label{eq:meantransp}
\end{equation}
over the $m$ iterations of the external mode performed during the current step of the internal mode. While $\fdd$ can build similarly, it is more efficient to derive it from the 2D momentum before and after the 2D iterations ($\Transp_0$ and $\Transp_1$).
\begin{equation}
    \begin{split}    
        \fdd 
        &= \dfrac{1}{m}\sum_{\text{2D iterations}} \eint{\phi\fparti{\Transp}{t} \jj}_{\trih} - \fdddh \\
        &= (\Transp_1 - (\Transp_0 + \Delta t \fdddh)) / \Delta t\label{eq:f2d:discrete}
    \end{split}
\end{equation}

\subsection{3D Diagnostic Equations}
This section focuses on the two main diagnostic equations of the 3D model: the continuity equation~\eqref{eq:conti3d} and the equation for the horizontal pressure gradient~\eqref{eq:hpg}.
\vspace{0.3cm}
\subsubsection{Continuity Equation}
The continuity equation~\eqref{eq:conti3d} is solved on the 3D mesh, which is slightly non-conformal. In general, a lateral interface between two prisms does not coincide exactly on both sides, creating an ambiguity in the integration domain for the lateral terms. To avoid this, all integrals are expressed in the parent element, and variations in the integration domain are represented through a discontinuous Jacobian across the lateral interfaces. 
\begin{equation}\label{eq:conti3d}
    \fparti{w}{z} = - \gradh \cdot \uuu 
\end{equation}

The discrete form of equation~\eqref{eq:conti3d} is based on the following weak formulation:
\[
    \begin{split}    
        \eeint{\phi \nz w^*}_{\gtb} - \eint{w\fparti{\phi}{z}}_{\prism} 
        &= \eint{\uuu \cdot \gradh\phi}_{\prism} - \eeint{\phi \nn_h \cdot \uuu^*}_{\gtb} \\
        &- \eeint{\phi \nn_h \cdot \uuu^*}_{\glat},
    \end{split}
\]
which, when expressed in the parent element, becomes
\[
    \begin{split}        
        \eeint{\phi \nz w^* \jtb}_{\gtbh} &- \eint{w\fparti{\phi}{z} \jz \jj}_{\prismh} \\
        =& \eint{\jz \uuu \cdot \gradh\phi \jj}_{\prismh} - \eeint{\jtb \phi \nn_h \cdot \uuu^*}_{\gtbh} \\
        &- \eeint{\phi \nn_h \cdot \uuu^*  \jz\jedge}_{\glath}.
    \end{split}
\]
Here, we use the tensor-product structure of the mesh, where the volume Jacobian satisfies $J = \jz \jj$ and the lateral Jacobian $\jlat = \jz\jedge$.

A key requirement of the continuity equation is its consistency with the free-surface equation~\eqref{eq:mass2d} in some parts of the temporal scheme. In practice, this means that the vertical integral of~\eqref{eq:conti3d}, combined with the impermeability condition at the ocean floor, must reproduce the free-surface equation~\eqref{eq:mass2d}. This property must also hold in the discrete formulation. To ensure this, the discrete continuity equation employs a linearized transport $\transp$, consistent with the averaged 2D linear transport $\bar{\Transp}$ from equation~\eqref{eq:meantransp}.

% however, since $\jz$ and $\uuu$ are both linear, $\transp$ is formally quadratic. 
In the continuous limit, we have $\transp = \jz \uuu$. In practice, however, $\transp$ is always projected onto the linear basis functions from $\jz \uuu$. When consistency is required, the 3D transport is build to match the mean 2D transport $\bar{\Transp}$ and denoted as $\bar{\transp}$.
\[
    \sum_\text{Vertical DOFs} \bar{\transp} = \bar{\Transp}
\]
Otherwise, $\transp$ is simply the projection of $\jz \uuu$ to the linear basis functions.
To express the continuity equation in terms of $\transp$, we replace $\jz\uuu$ by $\transp$ wherever possible, and use $\transp/\jz$ in place of $\uuu$ otherwise:
\[
    \begin{split}    
        \eeint{\phi \nz w^* \jtb}_{\gtbh} &- \eint{w\fparti{\phi}{z} \jz \jj}_{\prismh} \\
        =& \eint{\transp \cdot \gradh\phi \jj}_{\prismh} - \eeint{\phi \nn_h \cdot (\transp/\jz)^* \jtb}_{\gtbh} \\
        &- \eeint{\phi \nn_h \cdot \transp^* \jedge}_{\glath}.
    \end{split}
\]
This equation can be simplified by noting that
\begin{equation}\label{eq:simplification:whpg}
    \jtb = \abs{\jj/\nz} \qquad \text{and} \qquad \fparti{}{\hat{z}} = \dfrac{\partial z}{\partial \hat{z}}\fparti{}{z} = \jz \fparti{}{z},
\end{equation}
which gives
\[
    \begin{split}    
        \eeint{\phi \nzh w^* \jj}_{\gtbh} &- \eint{w\fparti{\phi}{\hat{z}}\jj}_{\prismh} \\
        =& \eint{\transp \cdot \gradh\phi \jj}_{\prismh} - \eeint{\phi \nn_h \cdot (\transp/\jz)^*\abs{\jj/\nz}}_{\gtbh} \\
        &- \eeint{\phi \nn_h \cdot \transp^* \jedge}_{\glath}.
    \end{split}
\]
We now define the interface variables explicitly:
\begin{itemize}
    \item In the term $\eeint{\phi \nzh w^* \jj}_{\gtbh}$, $w^*$ corresponds to the value of $w$ below the interface. Thus, at the top of an element, $w$ is taken from the interior, while at the bottom it is taken from the exterior. This choice is consistent with the vertical nature of the equation and the impermeability condition at the ocean floor.
    \item In the term $\eeint{\phi \nn_h \cdot (\transp/\jz)^*\abs{\jj/\nz}}_{\gtbh}$, the interface value is computed as the simple average $(\transp/\jz)^* = \bmean{\transp/\jz}$.
    \item The term $\eeint{\phi \nn_h \cdot \transp^* \jedge}_{\glath}$ requires particular care to ensure consistency with the boundary terms of equation~\eqref{eq:moment2d:discrete}. To achieve this, we define the lateral flux as
    \[
        \transp^* = \bmean{\transp} + \nn_h \bmean{\jz/H}\bmax{c}\bdiff{\eta}.
    \]
\end{itemize}
This yields the final discrete continuity equation:
\begin{equation}\label{eq:conti3d:discrete}
\begin{split}
    &\eeint{\phi w\inte \nzh \jj}_{\gtoph} + \eeint{\phi w\exte \nzh \jj}_{\gboth} - \eint{w\jj\fparti{\phi}{\hat{z}}}_{\prismh} \\
    &\quad= \eint{\transp \cdot \gradh \phi \jj}_{\prismh} - \eeint{\nn_h \cdot\bmean{\transp/\jz} \phi \abs{\jj/\nz}}_{\gtbh} \\
    &\quad- \eeint{\phi \left(\nn_h \cdot\bmean{\transp} + \bmean{\jz/H}\bmax{c}\bdiff{\eta}\right) \jedge}_{\glath}.
\end{split}
\end{equation}
As mentioned previously, $\transp$ should be understood as the linearized 3D horizontal transport $\jz\uuu$, i.e., a projection of the exact (second-degree) transport field onto the linear basis functions.
\vspace{0.3cm}
\subsubsection{Horizontal Pressure Gradient}\label{sec:hpg_detail}
The continuous equation for the horizontal pressure gradient $\hpg$ reads
\begin{equation}\label{eq:hpg}
    \fparti{\hpg}{z} = g\gradh \rho',
\end{equation}
with its corresponding weak form
\[
    \eeint{\phi \nz \hpg^*} - \eint{\hpg \fparti{\phi}{z}} = - g\eint{\gradh \phi \rho'} + g\eeint{\phi \nn_h \rho'{}^*},
\]
which, when expressed in the parent element, becomes
\[
    \begin{split}    
        \eeint{\phi \nz \hpg^* \jtb}_{\gtbh} &- \eint{\hpg \fpart{\phi}{z} \jj \jz}_{\prismh} \\
        =& - g\eint{\gradh \phi \rho' \jj \jz}_{\prismh} + g\eeint{\phi \nn_h \rho'{}^* \jtb}_{\gtbh} \\
        &+ g\eeint{\phi \nn \rho'{}^* \jz^* \jedge}_{\glath} \;.
    \end{split}
\]

On the interfaces, the discontinuous variables are defined as follows:
\begin{itemize}
    \item In the term $\eeint{\phi \nz \hpg^* \jtb}_{\gtbh}$, $\hpg^*$ refers to the value of $\hpg$ above the interface. Thus, at the top of an element we use $\hpg$ from the exterior, while at the bottom we use $\hpg$ from the interior. As in equation~\eqref{eq:conti3d:discrete}, this choice follows from the vertical nature of the equation and the boundary condition being applied at the ocean surface.
    \item In the term $\eeint{\phi \nn_h \rho'{}^* \jtb}_{\gtbh}$, $\rho'{}^*$ also refers to the value above the interface, for the same reason.
    \item In $\eeint{\phi \nn \rho'{}^* \jz^* \jedge}_{\glath}$, we discretize $\rho'{}^* \jz^*$ as $\bmean{\rho'}\bmean{\jz}$. We do not use $\bmean{\rho'\jz}$, since for constant density fields with slightly discontinuous interfaces, this alternative would spuriously modify the flow through artificial density gradients.
\end{itemize}
With these definitions and applying the simplifications from equation~\eqref{eq:simplification:whpg}, we rewrite the equation as
\[
    \begin{split}
        &\eeint{\phi \hpg\exte \nzh\jj}_{\gtoph} + \eeint{\phi \hpg\inte \nzh\jj}_{\gboth} - \eint{\hpg \fpart{\phi}{\hat{z}} \jj}_{\prismh} \\
        &\quad= - g\eint{\gradh \phi \rho' \jj \jz}_{\prismh} + g\eeint{\phi \nn_h \rho'{}\exte \abs{\jj/\nz}}_{\gtoph} \\
        &\quad+ g\eeint{\phi \nn_h \rho'{}\inte \abs{\jj/\nz}}_{\gboth} 
        + g\eeint{\phi \nn \bmean{\rho'} \bmean{\jz} \jedge}_{\glath} \;.
    \end{split}
\]
As in Section~\ref{sec:momentum:2d:discrete}, we integrate the right-hand side backwards, leading to
\[
\begin{split}
    &\eeint{\phi \hpg\exte \nzh\jj}_{\gtoph} + \eeint{\phi \hpg\inte \nzh\jj}_{\gboth} - \eint{\hpg \fpart{\phi}{\hat{z}} \jj}_{\prismh} \\
    &\quad= g\eint{\phi \gradh \rho' \jj \jz}_{\prismh} + g\eeint{\phi \nn_h (\rho'{}\exte - \rho'{}\inte) \abs{\jj/\nz}}_{\gtoph} \\
    &\quad+ g\eeint{\phi \nn (\bmean{\rho'}  - \rho'{}\inte) \bmean{\jz} \jedge}_{\glath} \;.
\end{split}
\]
The integral over the bottom of each element cancels out during this backward integration. Since $\nzh$ appears only on the top face, it simplifies to $\nzh = 1$. Finally, using $(\rho'{}\exte - \rho'{}\inte) = -2\bdiff{\rho'}$ and $(\bmean{\rho'}  - \rho'{}\inte) = -\bdiff{\rho'}$, we obtain the final discrete form of equation~\eqref{eq:hpg}:
\begin{equation}\label{eq:hpg:discrete}
    \begin{split}
        &\eeint{\phi \hpg\exte \jj}_{\gtoph} + \eeint{\phi \hpg\inte \jj}_{\gboth} - \eint{\hpg\fpart{\phi}{\hat{z}} \jj}_{\prismh} \\
        &\quad= g\eint{\phi \gradh \rho' \jj \jz}_{\prismh} - g\eeint{2\phi \nn_h \bdiff{\rho'} \abs{\jj/\nz}}_{\gtoph} \\
        &\quad- g\eeint{\phi \nn_h \bdiff{\rho'}\bmean{\jz}\jedge}_{\glath}.
    \end{split}
\end{equation}

\subsection{Prognostic Equations}
In the 2D equations, only the spatial components are discretized, resulting in a system of ordinary differential equations (ODEs) in time. This system can then be integrated using any of a wide range of implicit or explicit time-stepping methods. In contrast, the discretization of the 3D equations follows the temporal scheme introduced in \cite{ishimweMultiscaleIMEXSecond2025}, which couples the spatial and temporal discretizations while treating the vertical terms implicitly in time.

\begin{figure}[htpb]
    \setfigurenum{S1}
    \centering
    \includegraphics[]{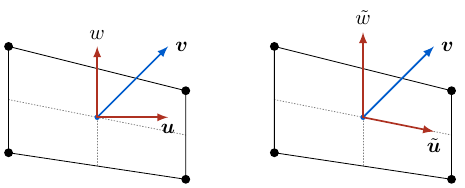}
    \caption{Left: original splitting, where the horizontal velocity $\uuu$ is treated explicitly and the vertical velocity $w$ implicitly. Right: modified splitting used in this work, where the mesh-aligned velocity component $\tilde{\uuu}$ is treated explicitly and the vertical component $\tilde{w}$ implicitly.}
    \label{fig:impl-expl-u-split}
\end{figure}

Compared to the original formulation in \cite{ishimweMultiscaleIMEXSecond2025}, this work adopts a slightly modified implicit/explicit splitting strategy. Rather than decomposing the 3D velocity field $\bm{v}$ into horizontal ($\uuu$) and vertical ($w$) components, we split it into a mesh-aligned component $\tilde{\uuu}$ and a vertical component $\tilde{w}$, as illustrated in Figure~\ref{fig:impl-expl-u-split}. The vertical component $\tilde{w}$ is treated implicitly, while the mesh-aligned component $\tilde{\uuu}$ is handled explicitly. Note that this \emph{does not} introduce any additional spatial error.

To explicitly define $\tilde{\uuu}$ in terms of $\uuu$, let us introduce the vector $\mm$ as
\[
    \mm = \fpartf{\zeta}{\xx} = \left(\fpartf{\zeta}{x}, \fpartf{\zeta}{y}, \fpartf{\zeta}{z}\right),
\]
where $\zeta$ denotes the vertical parametric coordinate inside the prism, ranging from $\zeta = -1$ (bottom) to $\zeta = 1$ (top). By construction, $\mm$ is aligned with the normal vector on the top and bottom faces of each element, and orthogonal to $\tilde{\uuu}$. We can then express
\[
    \tilde{\uuu} = (\uuu, -\mmh \cdot \uuu / \mmz) = (u, v, -\mmh \cdot \uuu / \mmz),
\]
while the corresponding vertical component is
\[
    \tilde{w} = w + \mmh \cdot \uuu / \mmz.
\]

The purpose of this decomposition is to minimize the explicit fluxes across the top and bottom faces. By construction, $\tilde{\uuu}$ is tangent to these faces, thereby eliminating explicit advective fluxes. To further reduce diffusive and viscous contributions, a similar splitting is applied to the diffusivity and viscosity tensors. Let $\tensor{D}$ denote either of these tensors. We decompose it as $\tensor{D} = \tensor{D}_i + \tensor{D}_e$, where $\tensor{D}_i$ and $\tensor{D}_e$ represent the implicit and explicit parts, respectively. The implicit component is defined as
\begin{equation}\label{eq:diff_split_impl_expl}
    \tensor{D}_i = \left(\dfrac{\mm \cdot \tensor{D} \cdot \mm}{\mmz^2}\right) \bm{e}_z \otimes \bm{e}_z 
    = \begin{bmatrix}
        0 & & \\
        & 0 & \\
        & & D_i
    \end{bmatrix}
    \;.
\end{equation}
Since $\tensor{D}_e = \tensor{D} - \tensor{D}_i$, this ensures that $\mm \cdot \tensor{D}_e \cdot \mm = 0$, and thus $\nn \cdot \tensor{D}_e \cdot \nn = 0$ on the top and bottom faces.

\vspace{0.3cm}
\subsubsection{Modified Continuity Equation}
Although $\tilde{w}$ can, in principle, be computed from $w$, $\uuu$, and $\mm$, it is more efficient and numerically stable to solve for $\tilde{w}$ directly. The governing equation for $\tilde{w}$ has the same form as equation~\eqref{eq:conti3d}:
\[
    \fparti{\tilde{w}}{z} = - \nabla \cdot \tilde{\uuu} \;.
\]
Following the same procedure as before, we express its weak form in the parent domain as
\[
    \begin{split}    
        &\eeint{\phi \nz \tilde{w}^* \jtb}_{\gtbh} - \eint{\tilde{w}\fparti{\phi}{z} \jz \jj}_{\prismh} \\
        &= \eint{\jz \tilde{\uuu} \cdot \nabla\phi \jj}_{\prismh} - \eeint{\jtb \phi \nn \cdot \tilde{\uuu}^*}_{\gtbh} \\
        &- \eeint{\phi \nn \cdot \tilde{\uuu}^*  \jz\jedge}_{\glath}.
    \end{split}
\]
Because $\tilde{\uuu}$ is aligned with the mesh, it is orthogonal to the normal vector on the top and bottom faces. Consequently, the integrals over $\gtbh$ vanish. On the lateral faces, the normal vector lies in the horizontal plane, so $\nn \cdot \tilde{\uuu} = \nn_h \cdot \uuu$, resulting in the same term as in the standard continuity equation. 

To simplify the volume term, we decompose the gradient operator as
\[
    \nabla = \gradhh + \mm \fpartf{}{\zeta},
\]
where
\[
    \gradhh \defeq \left(\fpartf{\xi}{x}\fpartf{}{\xi} + \fpartf{\eta}{x}\fpartf{}{\eta}, \fpartf{\xi}{y}\fpartf{}{\xi} + \fpartf{\eta}{y}\fpartf{}{\eta}, 0\right).
\]
The operator $\gradhh$ represents the horizontal gradient along iso-$\zeta$ slices of a prism. For 2D fields, or for fields that are vertically constant, $\gradhh$ reduces to $\gradh$. Applying this decomposition to $\nabla \phi$ yields
\[
    \nabla \phi = \gradhh \phi + \mm \fpartf{\phi}{\zeta} = \phiz \gradh \phih + \mm \fpartf{\phi}{\zeta}.
\]
Since $\tilde{\uuu}$ is perpendicular to $\mm$, the second term from $\tilde{\uuu} \cdot \nabla\phi$ vanishes, leaving $\tilde{\uuu} \cdot \gradhh \phi = \uuu \cdot \gradhh \phi$. Substituting this result, the integrals simplify to
\[
    \begin{split}    
        &\eeint{\phi \nz \tilde{w}^* \jtb}_{\gtbh} - \eint{\tilde{w}\fparti{\phi}{z} \jz \jj}_{\prismh} \\
        &\quad= \eint{\jj \jz \uuu \cdot \phiz\gradh\phih}_{\prismh} - \eeint{\phi \nn_h \cdot \uuu^*  \jz\jedge}_{\glath}.
    \end{split}
\]
From this point, the derivation proceeds identically to that leading to equation~\eqref{eq:conti3d:discrete}, resulting in the following discrete form for $\tilde{w}$:
\begin{equation}\label{eq:wtilde_discrete}
    \begin{split}
        &\eeint{\phi \tilde{w}\inte \nzh \jj}_{\gtoph} + \eeint{\phi \tilde{w}\exte \nzh \jj}_{\gboth} - \eint{\tilde{w}\jj\fparti{\phi}{\hat{z}}}_{\prismh} \\
        &\quad= \eint{\jj \transp \cdot \phiz \gradh \phih}_{\prismh} \\
        &\quad- \eeint{\phi \left(\nn_h \cdot\bmean{\transp} + \bmean{\jz/H}\bmax{c}\bdiff{\eta}\right) \jedge}_{\glath}.
    \end{split}
\end{equation}
The modified continuity equation thus provides a consistent definition of $\tilde{w}$ that is fully compatible with the implicit/explicit velocity decomposition introduced earlier. This expression of $\tilde{w}$ is then used in the implicit treatment of the vertical fluxes within the 3D momentum and tracer equations, discussed below.

\vspace{0.3cm}
\subsubsection{3D Momentum}\label{sec:3d_momentum_detail}
The discretization of equation \eqref{eq:moment3d:split} introduces additional complexity due to its two-way coupling with the external mode:
\begin{equation}\label{eq:moment3d:split}
    \begin{split}
        &\fparti{\uuu}{t} + \gradh \cdot (\uuu \otimes \uuu) + \fpart{w \uuu}{z} \\
        &\quad= \gradh \cdot (\kappa_h \gradh \uuu) +  \fpart{\kappa_v \fparti{\uuu}{z}}{z} \\
        &\quad- f \bm{e}_z \times \uuu - \dfrac{1}{\rho_0}\hpg  + S_2
    \end{split}
\end{equation}
To handle this coupled system, SLIM reformulates the discrete version of \eqref{eq:moment3d:split} as
\begin{equation}\label{eq:moment3d:discrete:simple}
    \dfrac{M_1 \uuu_1 - M_0 \uuu_0}{\Delta t} = \fddde(\uuu, \bar{\transp}, \hpg) + \dfrac{M_1\fdd}{H_1} + \fdddi(\uuu, \tilde{w}, \uuu_1) 
\end{equation}
when vertical quantities are treated implicitly, or
\begin{equation}\label{eq:moment3d:discrete:simple_expl}
    \dfrac{M_1 \uuu_1 - M_0 \uuu_0}{\Delta t} = \fddde(\uuu, \bar{\transp}, \hpg) + \dfrac{M_1\fdd}{H_1} + \fdddi(\uuu, \tilde{w}, \uuu) 
\end{equation}
when everything is explicit.
Here, $M_0$ and $M_1$ are the (time-varying) mass matrices at the beginning and end of the time step, respectively. $\uuu_0$ is the velocity field at the beginning of the step, and $\uuu_1$ is the unknown at its end. $\uuu$ represents the intermediate solution from the Runge–Kutta scheme. The transport $\bar{\transp}$ is derived from $\uuu$ and the mean transport $\bar{\Transp}$. The term $\fddde$ includes the horizontal advection and viscosity, the Coriolis force and the horizontal pressure gradient. $\fdddi$ accounts for the vertical advection and viscosity, and wind stresses and bottom friction as viscous boundary conditions. Finally, $\fdd$ represents the horizontal momentum change from the external mode, i.e. the change induced by $S_2$ and computed via \eqref{eq:f2d:discrete}. Since $\fdd$ is vertically integrated, a division by $H$ is required to obtain a velocity increment. In practice, $H$ is taken at the end of the step ($H_1$). The multiplication by the mass matrix ensures that the velocity increment is uniform vertically.

With this splitting, the interdependence between \eqref{eq:moment3d:split} and \eqref{eq:moment2d:split} becomes manageable. The 3D contribution to the 2D mode ($S_3$) corresponds to the vertical integral of \eqref{eq:moment3d:split} without $S_2$, which discretely translates to the sum over all vertical degrees of freedom of $\fddde$ and $\fdddi$:
\[
    \fdddh = \sum_\text{Vertical DOFs} (\fddde + \fdddi) \;.
\]
However, neglecting wind drag and bottom friction for the moment, $\fdddi$ integrates to zero over the vertical and is therefore not required for computing $\fdddh$. This allows equation \eqref{eq:moment2d:discrete} to be solved before $\fdddi$ is known. Wind drag and bottom friction must still be included in $\fdddh$. During explicit steps, their contribution can be added directly. For implicit steps, however, the bottom drag depends on the unknown velocity $\uuu_1$ and is therefore approximated using a prediction, rather than computed exactly. Wind drag, on the other hand, is treated explicitly. Since consistency is not relevant for $\fdddh$, $\fddde$ is evaluated with $\transp$ rather than $\bar{\transp}$.

With this strategy, the discretization of \eqref{eq:moment3d:split} proceeds as follows:
\begin{enumerate}
    \item Solve for $\hpg$ using \eqref{eq:hpg:discrete}.
    \item Compute $\transp$ by projecting $\jz \uuu$ to the linear basis functions.
    \item Compute $\fddde(\uuu, \transp, \hpg)$.
    \item Compute the wind drag $\bm{\tau}_s$ and compute or predict the bottom friction $\bm{\tau}_b^*$.
    \item Compute $\fdddh = \sum_\text{Vertical DOFs} \fddde + \bm{\tau}_s + \bm{\tau}_b^*$.
    \item Advance \eqref{eq:moment2d:discrete}, accumulating $\fdd$ and $\bar{\Transp}$ following \eqref{eq:f2d:discrete} and \eqref{eq:meantransp}.
    \item Update the vertical mesh coordinate ($z$) and velocity ($w_m$).
    \item Compute $\bar{\transp}$ from $\jz \uuu$ and $\bar{\Transp}$.
    \item Compute the modified vertical velocity $\tilde{w}$ from $\bar{\Transp}$.
    \item Solve equation \eqref{eq:moment3d:discrete:simple} or \eqref{eq:moment3d:discrete:simple_expl} for $\uuu_1$.
\end{enumerate}

The discrete form of $\fddde$ is
\begin{equation}\label{eq:moment3d:hor}
    \begin{split}
        \fddde(\uuu, \transp, \hpg)
        =& \eint{\jj \uuu (\transp \cdot \phiz \gradh \phih)}_{\prismh}  \\
        &- \eeint{\phi \uuu\upw \left(\nn_h \cdot\bmean{\transp} + \bmean{\jz/H}\bmax{c}\bdiff{\eta}\right) \jedge }_{\glath} \\
        &- \eint{J (\nabla \phi \cdot \tensor{\kappa}_e \cdot \nabla) \uuu}_{\prismh}  \\
        &+ \eeint{\phi \bmean{\jbnd (\nn  \cdot \tensor{\kappa}_e \cdot \nabla) \uuu}}_{\partial\prismh} \\
        &- \eeint{\sigma_3  \bmax{\nn \cdot \tensor{\kappa}_e \cdot \nn} \bmean{\jlat}\bdiff{\uuu}}_{\glath} \\
        &- \eint{J\phi f \bm{e}_z \times \uuu}_{\prismh} - \eint{J\phi \hpg/\rho_0}_{\prismh}.
    \end{split}
\end{equation}
The corresponding expression for $\fdddi$ is
\begin{equation}\label{eq:moment3d:vert}
    \begin{split}
        \fdddi(\uuu, \tilde{w}, \uuu_1)
        =& \eint{J \fpart{\phi}{z} (\tilde{w}-w_m)\uuu_1}_{\prismh} \\
        &- \eeint{\jj \nzh \phi \uuu_1\upw (\tilde{w}^\text{bot} - w_m)}_{\gtbh} \\
        &- \eint{J \fpart{\phi}{z} \kappa_i \fpart{\uuu_1}{z}}_{\prismh} \\
        &+ \eeint{\phi \bmean{\jtb \nz \kappa_i\partial_z\uuu_1}}_{\gtbh} \\
        &- \eeint{\sigma_3 \bmax{\kappa_i \nz^2} \jtb\bdiff{\uuu_1}}_{\gtbh}.
    \end{split}
\end{equation}

In both equations, the viscosity tensor $\tensor{\kappa}$ is parameterized using a horizontal viscosity $\kappa_h$ based on Smagorinsky’s formulation, and a vertical viscosity $\kappa_v$ provided by the turbulence closure model. The complete tensor $\tensor{\kappa}$ is split into explicit and implicit parts following \eqref{eq:diff_split_impl_expl}:

In the interface integrals, the upwind direction is determined from the advection velocity immediately adjacent to $\uuu\upw$. The mesh used in \eqref{eq:moment3d:hor} and \eqref{eq:moment3d:vert} always corresponds to that of $\uuu$, meaning $\uuu$ and the mesh come from the same Runge–Kutta stage. The interior penalty parameter \cite{riviereDiscontinuousGalerkinMethods2008}, \cite{rivierePenaltyfreeDiscontinuousGalerkin2014} $\sigma_3$ is defined as
\begin{equation}
    \sigma_d = \dfrac{N_0 (o+1)(o+d)}{2d \cdot \min(L\inte, L\exte)} \label{eq:iipm_factor}
\end{equation}
where $d=3$ is the problem dimension, $o=1$ is the polynomial degree of the basis functions, and $N_0=5$ is the average number of neighboring elements. $L$ approximates the effective length scale of an element. On horizontal interfaces, $L$ is taken as the average prism height, and on lateral faces, $L = A/l$ with $A$ the triangle area and $l$ the edge length. Although originally derived for simplices, this definition performs well in practice for prisms.

\vspace{0.3cm}
\subsubsection{Tracers}\label{sec:tracer3d_detail}
The discrete tracer equation is formulated on the same moving mesh as the momentum equation. Its discretization of advection and diffusion terms follows an almost identical structure, differing only by the absence of source terms such as the external mode, the wind, or the bottom friction. The resulting discrete form reads:
\begin{equation}\label{eq:tracer:discrete}
    \begin{split}
        &\dfrac{\eint{\phi T_1}_{\prism_1} - \eint{\phi T_0}_{\prism_0}}{\Delta t} = \eint{\jj T (\bar{\transp} \cdot \phiz \gradh \phih)}_{\prismh} \\
        &\qquad- \eeint{\phi T\upw \left(\nn_h \cdot\bmean{\bar{\transp}} + \bmean{\jz/H}\bmax{c}\bdiff{\eta}\right) \jedge }_{\glath} \\
        &\qquad- \eint{J (\nabla \phi \cdot \tensor{\nu}_e \cdot \nabla) T}_{\prismh} + \eeint{\phi \bmean{\jbnd (\nn  \cdot \tensor{\nu}_e \cdot \nabla) T}}_{\partial\prismh} \\
        &\qquad- \eeint{\sigma_3  \bmax{\nn \cdot \tensor{\nu}_e \cdot \nn} \bmean{\jlat}\bdiff{T}}_{\glath} \\
        &\qquad+ \eint{J \fpart{\phi}{z} (\tilde{w}-w_m)T_1}_{\prismh} - \eeint{\jj \nzh \phi T_1\upw (\tilde{w}^\text{bot} - w_m)}_{\gtbh} \\
        &\qquad- \eint{J \fpart{\phi}{z} \nu_i \fpart{T_1}{z}}_{\prismh} + \eeint{\phi \bmean{\jtb \nz \nu_i\partial_zT_1}}_{\gtbh} \\
        &\qquad- \eeint{\sigma_3 \bmax{\nu_i \nz^2} \jtb\bdiff{T_1}}_{\gtbh} \;.
    \end{split}
\end{equation}
Although the corrected transport $\bar{\transp}$ is strictly required only for tracer advection, it is also applied to the advection of momentum. Since $\bar{\transp}$ must be computed for tracers in any case, using it consistently for velocity ensures a unified and coherent treatment of advection with acceptable additional costs.

\printbibliography